\documentclass[journal, 10pt]{IEEEtran}
\usepackage[colorlinks,
            linkcolor=blue,
            anchorcolor=red,
            citecolor=red]{hyperref}

\usepackage{booktabs}

\usepackage{diagbox} 
\usepackage{graphicx}
\usepackage{epstopdf}
\usepackage{psfrag}
\usepackage{subfigure}
\usepackage{url}
\usepackage{stfloats}
\usepackage{amsfonts,amssymb,amsmath,bm,paralist,theorem,cite,ifthen,color,nccmath}
\usepackage{caption}
\usepackage{calc}
\usepackage{enumerate}
\usepackage{multirow}
\usepackage{makecell}
\usepackage[linesnumbered,ruled]{algorithm2e}
\usepackage{setspace}
\usepackage{array}
\usepackage{float}
\usepackage{soul}
\usepackage{titlesec}

\graphicspath{{Figures/}}
\newcommand\Ns{\ensuremath{ N_{\rm s} }}
\newcommand\Nt{\ensuremath{ N_{\rm t} }}
\newcommand\Nr{\ensuremath{ N_{\rm r} }}
\newcommand\Kcl{\ensuremath{\mathcal{K}}}
\newcommand\ssb{\ensuremath{ \mathbf{s} }}
\newcommand\Cs{\ensuremath{{\mathbb{C}}}}
\newcommand\Wb{\ensuremath{ \mathbf{W} }}
\newcommand\wb{\ensuremath{ \mathbf{w} }}

\newcommand\rmc{\ensuremath{ \mathrm{c} }}
\newcommand\rms{\ensuremath{ \mathrm{s} }}
\newcommand\rmf{\ensuremath{ \mathrm{F} }}

\newcommand\rmtr{\ensuremath{ \mathrm{tr} }}
\newcommand\Rx{\mathbf W\mathbf W^\H}

\def\BibTeX{{\rm B\kern-.05em{\sc i\kern-.025em b}\kern-.08em
T\kern-.1667em\lower.7ex\hbox{E}\kern-.125emX}}

\newtheorem{lemma}{Lemma}
\newtheorem{theorem}{Theorem}
\newtheorem{proposition}{Proposition}

\SetKwInput{Input}{input}
\SetKwInput{Output}{output}
\newtheorem{definition}{Definition}
\usepackage{array}
\newcolumntype{L}[1]{>{\raggedright\let\newline\\\arraybackslash\hspace{0pt}}m{#1}}
\newcolumntype{C}[1]{>{\centering\let\newline\\\arraybackslash\hspace{0pt}}m{#1}}
\newcolumntype{R}[1]{>{\raggedleft\let\newline\\\arraybackslash\hspace{0pt}}m{#1}}

\newtheorem{remark}{Remark}

\IEEEoverridecommandlockouts
\def\specialpapernotice#1{\if@confmode%
	\def\@specialpapernotice{{\sublargesize\textit{#1}\vspace*{1em}}}%
	\else%
	\def\@specialpapernotice{{\\*[1.5ex]\sublargesize\textit{#1}}\vspace*{-2ex}}%
	\fi}

\makeatletter\patchcmd{\@makecaption}{\scshape}{}{}{}

\usepackage[all=normal,paragraphs=tight,floats=normal,mathspacing=normal,wordspacing=tight,charwidths=tight,mathdisplays=normal,leading=normal]{savetrees}

\newcommand{\T}{{\scriptscriptstyle\mathsf{T}}}
\renewcommand{\H}{{\scriptscriptstyle\mathsf{H}}}

\allowdisplaybreaks
\begin{document}
\setlength{\abovedisplayskip}{3.5pt}
\setlength{\belowdisplayskip}{3.5pt}

\titlespacing{\section}{-0.64 cm}{4pt}{2pt}
\titlespacing{\subsection}{0 cm}{4pt}{2pt}

\title{Optimal ISAC Beamforming Structure and Efficient Algorithms for Sum Rate and CRLB Balancing}
\author{Tianyu Fang, \IEEEmembership{Student Member, IEEE}, Mengyuan Ma, \IEEEmembership{Student Member, IEEE}, Markku Juntti, \IEEEmembership{Fellow, IEEE}, Nir Shlezinger, \IEEEmembership{Senior Member, IEEE}, A.~Lee~Swindlehurst, \IEEEmembership{Fellow, IEEE}, and Nhan Thanh Nguyen, \IEEEmembership{Member, IEEE}
\vspace{-0.8cm}
\thanks{
A part of this work has been accepted for presentation at the IEEE Int. Conf. Acoustics, Speech, and Signal Process. (ICASSP), 2025 \cite{fang2025low}.
\par T. Fang, M. Ma, M. Juntti, and N. T. Nguyen are with the Centre for Wireless Communications, University of Oulu, 90014 Oulu, Finland (e-mail:
\{tianyu.fang, mengyuan.ma, markku.juntti, nhan.nguyen\}@oulu.fi). N. Shlezinger is with School of ECE, Ben-Gurion University of the Negev, Beer-Sheva 84105, Israel (email: nirshl@bgu.ac.il). A. L. Swindlehurst are with the Center for Pervasive Communications \& Computing, University of California, Irvine, CA 92697, USA (email: swindle@uci.edu).}}


\maketitle

\begin{abstract}
Integrated sensing and communications (ISAC) has emerged as a promising paradigm to unify wireless communications and radar sensing, enabling efficient spectrum and hardware utilization. A core challenge with realizing the gains of ISAC stems from the unique challenges of dual purpose beamforming design due to the highly non-convex nature of key performance metrics such as sum rate for communications and the Cramér–Rao lower bound (CRLB) for sensing. In this paper, we propose a low-complexity structured approach to ISAC beamforming optimization to simultaneously enhance spectral efficiency and estimation accuracy. Specifically, we develop a successive convex approximation (SCA) based algorithm which transforms the original non-convex problem into a sequence of  convex subproblems ensuring convergence to a locally optimal solution. Furthermore, leveraging the proposed SCA framework and the Lagrange duality, we derive the optimal beamforming structure for CRLB optimization in ISAC systems. Our findings characterize the reduction in radar streams one can employ without affecting performance. This enables a dimensionality reduction that enhances computational efficiency. Numerical simulations validate that our approach achieves comparable or superior performance to the considered benchmarks while requiring much lower computational costs.

\end{abstract}

\begin{IEEEkeywords}
 Integrated sensing and communications, Cramér-Rao lower bound, successive convex approximation.
\end{IEEEkeywords}

\section{Introduction}\label{Sec:intro}
\par Future wireless communications technologies are anticipated to meet increasingly demanding objectives to enable a wide range of innovative applications. These applications, including autonomous vehicles, extended reality, smart factories, and advanced healthcare systems, require not only ultra-reliable and low-latency communications but also precise sensing capabilities for environment perception, localization, and tracking \cite{dong2025communication}. A major challenge lies in the limited availability of spectrum, as both communications and sensing  traditionally operate in separate frequency bands with isolated infrastructures. This separation results in inefficient spectrum utilization and high hardware costs, further compounded by the need to simultaneously support diverse use cases. Within this landscape, integrated sensing and communications (ISAC) has emerged as an attractive paradigm that unifies wireless communications and radar sensing functionalities \cite{liu2022integrated}. 

ISAC offers significant advantages over conventional technologies that treat communications and sensing as separate systems. By sharing the same radio-frequency signals, ISAC improves spectrum efficiency, reduces hardware and infrastructure costs, and enhances energy efficiency~\cite{ma2020joint}. Unlike traditional systems where communications and radar sensing  are typically designed and deployed independently, ISAC aims to achieve dual functionality with a unified system framework through shared spectrum, hardware, and joint signal processing \cite{zhang2021overview}. The need to have both functionalities co-exist while sharing resources, gives rise to the need for dedicated beamforming techniques that account for both functionalities while being complexity-aware~\cite{zhang2025low}.

\subsection{Related Works}
A substantial body of research has focused on beamforming optimization to enable attaining the aforementioned benefits in multiuser multi-antenna ISAC systems. Common metrics for the communications functionality are the achievable (sum) rate and signal-to-interference-plus-noise ratio (SINR). For sensing performance,  various metrics can be used, such as proximity to a desired beampattern~\cite{liu2018mu, liu2020joint, nhan2024joint,nhan2023deep, nhan2023multiuser}, signal-to-clutter-plus-noise ratio (SCNR)~\cite{wang2024lowcomplexity,fang2024beamforming, zhang2023edge, wang2024starris,ma2024model}, Cramér--Rao lower bound (CRLB) of the target parameters~\cite{liu2021cramer, song2023intelligent,xiong2023fundamental,zhou2024cram, ren2022fundamental}, and mutual information \cite{peng2024mutual,ouyang2022performance}.
For communications, the sum rate (SR) is often preferred, being is a fundamental measure of the overall multiuser communications network. For sensing, the CRLB is recognized as a fundamental metric that provides a theoretical limit on parameter estimation accuracy \cite{xiong2023fundamental, ren2022fundamental}. 
While the  SR and CRLB are highly relevant for assessing ISAC beamforming,  utilizing them as objectives for  beamformer optimization poses signicant challenges due to their being highly non-convex functions of the beamforming vectors. 

Various solutions have been proposed to address CRLB-based ISAC beamforming optimization problems, ranging from globally optimal to suboptimal. Liu \textit{et al.}~\cite{liu2021cramer} considered CRLB minimization subject to communications quality of service and transmit power constraints, and obtained a globally optimal solution using semi-definite relaxation (SDR). Recently, Tang \textit{et al.}~\cite{tang2025dual} extended the work of \cite{liu2021cramer} from a single-target to a multi-target scenario, proposed an upper bound to approximate the CRLB minimization problem, and developed an efficient algorithm based on the alternating direction method of multipliers (ADMM) framework to solve the approximated problem. However, these methods face inherent challenges when applied to beamforming designs that involve communications sum rate (SR) optimization. To tackle such challenges, Chen \textit{et al.}~\cite{chen2024transmitter} and Zhu \textit{et al.}~\cite{zhu2023integrated} integrated SDR with techniques such as successive convex approximation (SCA) or weighted minimum mean square error (WMMSE). Their approaches aim to achieve a locally optimal solution to the problem of balancing the SR and CRLB performance using a weighted objective that is subject to a transmit power constraint. Recently, a globally optimal beamforming optimization algorithm employing SDR and the branch and bound (BB) method was proposed in \cite{wu2024global} to optimize the weighted sum of the SR and CRLB. Although offering superior performance, the use of BB requires a high complexity that increases exponentially with the number of communications users. Fractional programming (FP) is another widely employed technique for CRLB optimization \cite{zhu2023integrated,Guo2023FP-ADMM,Liu2024SNR, chen2024fast}. For instance, in \cite{Guo2023FP-ADMM}, auxiliary variables are introduced to transform the fractional CRLB into a polynomial form, enabling a multivariable reformulation of the original problem that is solved using the ADMM. 

Beyond traditional model-based optimization approaches, deep learning has emerged as a useful tool beamforming design. For instance, a convolutional long short-term memory network leveraging historical channel data has been developed for predictive beamforming in ISAC-based vehicle-to-infrastructure networks in \cite{liu2022learning}. An alternative approach that combines deep learning with conventional optimization is deep unfolding~\cite{shlezinger2022model}, which was shown to facilitate iteration-limited optimization of ISAC beamforming~\cite{nhan2024joint, zhang2025low,zhang2024joint}.

While the above approaches provide effective numerical solutions to joint ISAC beamforming designs, they fail to offer fundamental insights into the structure of the optimal beamformers. As a result, they face challenges in computational complexity, especially in large-scale networks. To develop an efficient algorithm with lower computational complexity for ISAC beamforming, one promising approach is to identify and exploit the optimal beamforming structure (OBS) of the original problem. The OBS for multi-user communications-only systems was established in \cite{emil2014optimal}, where the minimum mean square error (MMSE) filter is proven to be the optimal choice for both transmit and receive beamforming using an SINR-based criterion. For sensing-only systems, the structure of the transmit beamforming covariance matrix has been characterized in \cite{Li2008range} for CRLB-based optimization, but the exact beamforming structure remains unknown. The goal of this paper is to address this problem for beamforming design in CRLB-based ISAC systems.

\subsection{Contributions}

We focus on the fundamental problem of maximizing the SR while minimizing the CRLB in a multi-user, multi-antenna monostatic ISAC system under a transmit power constraint, balancing these key communications and sensing metrics. Most existing CRLB-based optimization works focus on either single-target scenarios \cite{liu2021cramer, song2023intelligent,xiong2023fundamental,zhou2024cram, ren2022fundamental, wu2024global, Guo2023FP-ADMM,Liu2024SNR, chen2024fast} or multi-target cases with a single parameter \cite{tang2025dual, chen2024transmitter, zhu2023integrated}, typically assuming a uniform linear array (ULA). In contrast, we establish a general ISAC system model that supports simultaneous communications with multiple users and estimation of multiple parameters for multiple targets using a uniform planar array (UPA). Unlike traditional MIMO radar~\cite{stoica2007probing}, our model allows the number of radar signals to differ from the number of array elements for complexity reduction, allowing to characterize its impact on performance.
While the formulated problem is complex and inherently non-convex, our proposed approach identifies the OBS of CRLB-based beamforming optimization and is thus able to provide a flexible and low-complexity solution for CRLB-based optimization in ISAC systems. 

The key contributions of this paper are as follows:
\begin{itemize}
    \item We establish a generic monostatic ISAC system model that supports simultaneous communications with multiple users and detection of multiple targets.  We further derive a general CRLB-based radar sensing metric capable of estimating the azimuth and elevation angles of targets relative to the BS, as well as their complex radar cross-sections. This formulation encompasses existing CRLB metrics for single-target sensing or limited-parameter detection as special cases. Based on the derived CRLB and SR metrics, we formulate a highly challenging non-convex problem to achieve a tradeoff between communications and sensing performance. 
    \item We propose a unified framework based on SCA and the identified full power property to handle the non-convexity arising from both the communications and sensing objectives. A two-layer SCA method is proposed to transform the original highly non-convex problem into a sequence of simple sphere projection problems, whose optimal solutions are easy to obtain. We further provide a rigorous proof of convergence for the proposed method illustrating that the proposed algorithm reaches a locally optimal solution for the weighted sum of SR and CRLB maximization problem.
    \item Leveraging the proposed SCA method and Lagrange duality, we derive the OBS for CRLB-based ISAC systems. By exploiting the inherent low-dimensional beamforming structure, we reformulate the original problem into a lower-dimensional optimization with significantly fewer optimization variables.   The reformulation preserves the structure of the original problem but enables a highly efficient solution. In doing so, we rigorously characterize the allowable reduction in radar signal streams without compromising sensing performance, making the proposed beamforming design framework  particularly appealing for large-scale ISAC systems.
    \item We perform extensive simulations to demonstrate that the proposed algorithms achieve the same or better performances compared to benchmark optimization techniques, including SDR-based and FP-based algorithms, while significantly reducing the computational complexity.  
\end{itemize}

\textit{Organization:}  The rest of this paper is organized as follows. Section \ref{System model} introduces the system model and problem formulation. Section \ref{sec: algorithm} presents the proposed SCA method along with a thorough analysis of its optimality and convergence. Section \ref{Sec: OBS} derives the main results on the structure of the optimal beamforming solution. Simulation results and corresponding discussions are provided in Section \ref{Sec:simulation} to demonstrate the effectiveness of the proposed algorithms. Finally, Section \ref{Sec:concu} concludes the paper and discusses potential directions for future work.

\textit{Notation:} Vectors and matrices are represented by lowercase and uppercase bold letters, respectively. The space of complex- and real-valued numbers is respectively denoted by $\mathbb{C}$ and $\mathbb{R}$, while $\mathbb{R}_{++}$ represents the set of positive real numbers. The set of positive definite matrices is represented by $\mathbb S_{++}$. The expectation of a random variable  is denoted as $\mathbb{E}\{ \cdot \}$. The magnitude of a complex number is written as $|\cdot|$. The circularly symmetric complex Gaussian  distribution with zero mean and variance $\sigma^2$ is represented as $\mathcal{CN}(0,\sigma^2)$. The transpose and conjugate transpose operators are denoted by $(\cdot)^\T$ and $(\cdot)^\H$, respectively, while $\mathrm{diag}{\{\mathbf{y}\}}$ denotes a diagonal matrix whose main diagonal entries are elements of vector $\mathbf{y}$.

\section{Preliminaries}\label{System model}

\subsection{Signal Model}
\begin{figure}[t]
\center
\includegraphics[width=0.5\textwidth]{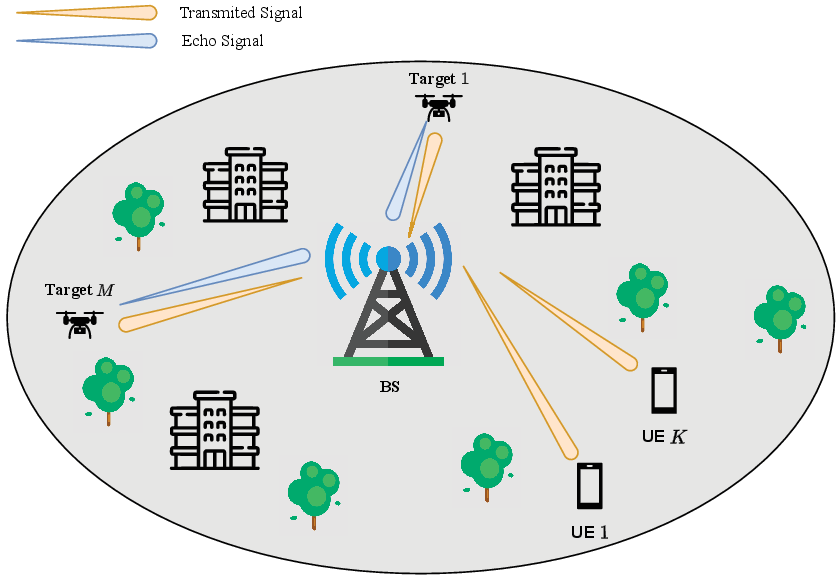}
    \vspace{-2mm}
    \caption{Illustration of the system model.}
    \label{fig:system model}
\end{figure}
We consider a downlink monostatic ISAC system as shown in Fig.~\ref{fig:system model}, where a base station (BS) is equipped with $ \Nt $ transmit antennas and $ \Nr $ antennas for radar reception, arranged as separate UPAs with half-wavelength spacing. 
The BS simultaneously transmits data signals to $K$ single-antenna communications users, indexed by the set $\Kcl\triangleq\{1,\ldots,K\}$, while also utilizing these signals to estimate the parameters of $M$ point-like  targets indexed by the set $\mathcal M\triangleq\{1,\ldots,M\}$, over $L$ time slots, indexed by $\mathcal L\triangleq\{1,\ldots,L\}$. 

\subsubsection{Transmitted Signal Model}
Let $ \ssb_{\mathrm c}[l]=[s_{\mathrm c 1}[l],\ldots, s_{\mathrm c K}[l]]^\T \in \Cs^{K\times 1} $ represent the vector of symbols to be transmitted to the $K$ users in the $l$-th time slot and let $ \Wb_{\mathrm c}\triangleq[\wb_{\mathrm c 1},\ldots,\wb_{\mathrm c  K}]\in\Cs^{\Nt\times K} $ denote the corresponding communications beamforming matrix at the BS. Similarly, let $\mathbf s_{\rms}[l] \in \Cs^{\Ns\times 1}$ be a vector of $\Ns$ signals at time slot $l$ devoted to radar sensing, which will be multiplied by the radar beamforming matrix $\mathbf W_\rms\in\mathbb C^{\Nt\times\Ns}$. While the common practice in MIMO radar~\cite{stoica2007probing}, also suggested for ISAC in \cite{liu2020joint}, is to set $\Ns=\Nt$ to exploit all available degrees of freedom (DoF) for sensing, this increases the complexity of the ISAC design and interference for communications. We will examine later how the choice of $\Ns$ affects the overall system performance. The complete transmit signal in the $l$-th time slot is given by $ \mathbf x[l]= \Wb_\rmc \ssb_\rmc[l]+\mathbf W_\rms \ssb_\rms[l]$, $\forall l\in\mathcal L$. It is assumed that the signal streams have unit power and are independent of each other, leading to $\mathbb E \{\ssb_\rmc[l]\ssb_\rmc^\H[l]\}=\mathbf I_K, \mathbb E \{\ssb_\rms[l]\ssb_\rms^\H[l]\}=\mathbf I_{\Ns}$, and $\mathbb E\{\ssb_\rmc[l]\ssb_\rms^\H[l]\}=\mathbf 0_{K\times\Ns}  $ \cite{liu2021cramer}.

\subsubsection{Communications Model}
The signal received by communications user $ k\in\mathcal K $ at time slot $l\in\mathcal L$ is given by
\begin{equation}
    y_{k}[l]\!= \!\mathbf h_k^\H\mathbf w_{\rmc k}s_{\rmc k}[l]\!+\!\sum_{j\neq k}^K \mathbf h^\H_k\mathbf w_js_{\rmc j}[l]
    \!+\!\mathbf h_k^\H\mathbf W_\rms\mathbf s_\rms[l]\!+\! n_{\rmc k}[l], 
\end{equation}
where $ \mathbf h_k\in\mathbb C^{\Nt\times 1} $ is the channel vector between the BS and user $ k $, and $ n_{\rmc k}[l]\sim\mathcal{CN}(0,\sigma_{\mathrm{c} k}^2)  $ is additive white Gaussian noise. The communications channels are assumed to be perfectly known at the BS and remain invariant throughout the entire transmission block. Consequently, the achievable rate in nats/s/Hz for each user $ k\in\mathcal K $ is given by
\begin{equation}\label{rate}
	R_{ k}=\log\!\left(\!\!1\!+\!\frac{|\mathbf h_k^\H\mathbf w_{\rmc k}|^2}{\sum_{j\neq k}^K|\mathbf h^\H_k\mathbf w_{\rmc j}|^2+\|\mathbf h_k^\H\mathbf W_\rms\|_\rmf^2+\sigma_{\mathrm{c} k}^2}\right), 
\end{equation} 
where $\|\cdot\|_\rmf$ denotes the Frobenius norm. The SR of all the communications users, given as $\sum_{k=1}^K R_k$, is used as the metric to evaluate the system's communications performance.

\subsubsection{Radar Model}
At time slot $l$, the received echo signal at the BS is given by
\begin{equation}\label{sensing signal matrix}
    \mathbf y_{\mathrm{s}}[l]=\sum_{m=1}^M \alpha_m \mathbf b(\theta_m,\phi_m)\mathbf a^\H(\theta_m,\phi_m)\mathbf x[l]+\mathbf n_{\mathrm{s}}[l],
\end{equation}
where $\alpha_m$ refers to the radar cross-section for target $m$, $\mathbf n_{\mathrm{s}}[l] \sim \mathcal{CN}(0,\sigma_s^2\mathbf{I})$ is noise at the receiver, $ \mathbf b(\cdot)$ and $\mathbf a(\cdot) $ are the receive and transmit array steering vectors for sensing and communications, respectively, with $\theta_m\in[-\pi,\pi]$ and $\phi_m\in[-\pi/2,\pi/2]$ representing the azimuth and elevation angles of the target relative to the BS. 

We use the CRLB associated with $\theta_m$ and $\phi_m$ as the sensing performance metric, which is obtained from the inverse of the Fisher information matrix (FIM). To derive the FIM, we begin by rewriting \eqref{sensing signal matrix} as
\begin{equation}
     \mathbf y_{\mathrm{s}}[l]=\mathbf B\mathbf U\mathbf A^\H\mathbf x[l]+\mathbf n_{\mathrm{s}}[l],
\end{equation}
where 
\begin{align}
\mathbf B &\triangleq[\mathbf b(\theta_1,\phi_1),\ldots,\mathbf b(\theta_M,\phi_M)] \\
\mathbf U &\triangleq\mathrm{diag}([\alpha_1,\ldots,\alpha_M]) \\
\mathbf A &\triangleq[\mathbf a(\theta_1,\phi_1),\ldots,\mathbf a(\theta_M,\phi_M)] .
\end{align}
Define $\bm \theta = [\theta_1,\cdots,\theta_M]^\T$, $\bm \phi= [\phi_1,\cdots,\phi_M]^\T$, and $\bm \alpha = [\alpha_1,\cdots,\alpha_M]^\T$, and 
let $\bm\omega=[\bm\theta,\bm\phi,\Re\{\bm\alpha\},\Im\{\bm\alpha\}]^\T\in\mathbb R^{4M\times 1}$ denote the vector of unknown parameters. We can now characterize the FIM as stated in the following lemma: 
\begin{lemma}
\label{lem:FIM}
    The FIM for estimating the parameters in $\bm\omega$ is given by the block matrix 
    \begin{align}\label{eq:CRLB}
		\mathbf F\!=\!  \frac{2L}{\sigma_s^2}
		\begin{bmatrix}
			 \Re\{\mathbf F_{11}\}   & \Re\{\mathbf F_{12}\}&\Re\{\mathbf F_{13}\}& -\Im\{\mathbf F_{13}\}\\
			\Re\{\mathbf F_{12}^\T\} &\Re\{\mathbf F_{22}\} & \Re\{\mathbf F_{23}\}& -\Im\{\mathbf F_{23}\}\\
    \Re\{\mathbf F_{13}^\T\} & \Re\{\mathbf F_{23}^\T\} & \Re\{\mathbf F_{33}\}&-\Im\{\mathbf F_{33}\}\\
    -\Im\{\mathbf F_{13}^\T\} &-\Im\{ \mathbf F_{23}^\T\} &-\Im\{ \mathbf F_{33}^\T\}& \Re\{\mathbf F_{33}\}
		\end{bmatrix},
\end{align}
whose block entries  are given by \eqref{FIM:blocks}, shown at the top of the next page, where $\mathbf R_x =\mathbf W_\rmc\mathbf W_\rmc^\H+\mathbf W_\rms\mathbf W_\rms^\H$, $\odot$ denotes the Hadamard product, and 
\begin{align}
        &\dot{\mathbf A}_{\bm\theta}\triangleq\left[\frac{\partial\mathbf a(\theta_1,\phi_1)}{\partial \theta_1},\ldots,\frac{\partial\mathbf a(\theta_M,\phi_M)}{\partial \theta_M}\right],\nonumber\\
        &\dot{\mathbf B}_{\bm\theta}\triangleq\left[\frac{\partial\mathbf b(\theta_1,\phi_1)}{\partial \theta_1},\ldots,\frac{\partial\mathbf b(\theta_M,\phi_M)}{\partial \theta_M}\right],\nonumber\\
        &\dot{\mathbf A}_{\bm\phi}\triangleq\left[\frac{\partial\mathbf a(\theta_1,\phi_1)}{\partial \phi_1},\ldots,\frac{\partial\mathbf a(\theta_M,\phi_M)}{\partial \phi_M}\right],\nonumber\\
        &\dot{\mathbf B}_{\bm\phi}\triangleq\left[\frac{\partial\mathbf b(\theta_1,\phi_1)}{\partial \phi_1},\ldots,\frac{\partial\mathbf b(\theta_M,\phi_M)}{\partial \phi_M}\right].\nonumber
    \end{align}
\end{lemma}

\begin{figure*}[ht]
    \begin{align}\label{FIM:blocks}
		&\mathbf F_{11} \triangleq\left(\mathbf U\mathbf A^\H\mathbf R_{\mathbf x}\mathbf A\mathbf U^\H\right)^\T \!\odot \!(\dot{\mathbf B}_{\bm\theta}^\H\dot{\mathbf B}_{\bm\theta})\!+\!(\mathbf U\mathbf A^\H\mathbf R_{\mathbf x}\dot{\mathbf A}_{\bm\theta}\mathbf U^\H)^\T\!\odot\!({\mathbf B}^\H\dot{\mathbf B}_{\bm\theta})\!+\!(\mathbf U\dot{\mathbf A}_{\bm\theta}^\H\mathbf R_{\mathbf x}\mathbf A\mathbf U^\H)^\T\!\odot\!(\dot{\mathbf B}_{\bm\theta}^\H{\mathbf B})\!+\!(\mathbf U\dot{\mathbf A}_{\bm\theta}^\H\mathbf R_{\mathbf x}\dot{\mathbf A}_{\bm\theta}\mathbf U^\H)^\T\!\odot\!({\mathbf B}^\H{\mathbf B})\nonumber \\
  &\mathbf F_{12} \triangleq(\mathbf U\mathbf A^\H\mathbf R_{\mathbf x}\mathbf A\mathbf U^\H)^\T\!\odot\!(\dot{\mathbf B}_{\bm\theta}^\H\dot{\mathbf B}_{\bm\phi})\!+\!(\mathbf U\mathbf A^\H\mathbf R_{\mathbf x}\dot{\mathbf A}_{\bm\theta}\mathbf U^\H)^\T\!\odot\!({\mathbf B}^\H\dot{\mathbf B}_{\bm\phi})\! +\!(\mathbf U\dot{\mathbf A}_{\bm\phi}^\H\mathbf R_{\mathbf x}\mathbf A\mathbf U^\H)^\T\!\odot\!(\dot{\mathbf B}_{\bm\theta}^\H{\mathbf B})\!+\!(\mathbf U\dot{\mathbf A}_{\bm\phi}^\H\mathbf R_{\mathbf x}\dot{\mathbf A}_{\bm\theta}\mathbf U^\H)^\T\!\odot\!({\mathbf B}^\H{\mathbf B}) \nonumber \\
& \mathbf F_{22} \triangleq(\mathbf U\mathbf A^\H\mathbf R_{\mathbf x}\mathbf A\mathbf U^\H)^\T\!\odot\!(\dot{\mathbf B}_{\bm\phi}^\H\dot{\mathbf B}_{\bm\phi})\!+\!(\mathbf U\mathbf A^\H\mathbf R_{\mathbf x}\dot{\mathbf A}_{\bm\phi}\mathbf U^\H)^\T\!\odot\!({\mathbf B}^\H\dot{\mathbf B}_{\bm\phi})\! +\!(\mathbf U\dot{\mathbf A}_{\bm\phi}^\H\mathbf R_{\mathbf x}\mathbf A\mathbf U^\H)^\T\!\odot\!(\dot{\mathbf B}_{\bm\phi}^\H{\mathbf B})\!+\!(\mathbf U\dot{\mathbf A}_{\bm\phi}^\H\mathbf R_{\mathbf x}\dot{\mathbf A}_{\bm\phi}\mathbf U^\H)^\T\!\odot\!({\mathbf B}^\H{\mathbf B}) \nonumber \\
		&\mathbf F_{13} \triangleq(\mathbf A^\H\mathbf R_{\mathbf x}\mathbf A\mathbf U^\H)^\T\odot(\dot{\mathbf B}_{\bm\theta}^\H\mathbf B) +(\mathbf A^\H\mathbf R_{\mathbf x}\dot{\mathbf A}_{\bm\theta}\mathbf U^\H)^\T\odot({\mathbf B}^\H\mathbf B) \nonumber  \\
  &\mathbf F_{23} \triangleq  (\mathbf A^\H\mathbf R_{\mathbf x}\mathbf A\mathbf U^\H)^\T\odot({\mathbf B}_{\bm\phi}^\H\mathbf B)  +(\mathbf A^\H\mathbf R_{\mathbf x}\dot{\mathbf A}_{\bm\phi}\mathbf U^\H)^\T\odot({\mathbf B}^\H\mathbf B) \nonumber \\
		&\mathbf F_{33}\triangleq(\mathbf A^\H\mathbf R_{\mathbf x}\mathbf A )^\T\odot(\mathbf B^\H\mathbf B),  
	\end{align}
\hrulefill 
\end{figure*}

\begin{IEEEproof}
    The lemma follows by expressing the FIM in terms of the noise-free signal $\mathbf v_\rms[l]= \mathbf y_{\mathrm{s}}[l]-\mathbf n_{\mathrm{s}}[l]=\mathbf B\mathbf U\mathbf A^\H\mathbf x[l]$ as \cite{bekkerman2006target}:
\begin{equation}\label{FIM:original}
       \hspace{-1mm}\mathbf F=2\Re \left\{\sum_{l=1}^L\frac{\partial \mathbf v_{\mathrm{s}}^\H[l]}{\partial \bm\omega}\mathbf R_{\mathbf n_{\mathrm s}}^{-1}\frac{\partial \mathbf v_{\mathrm{s}}[l]}{\partial \bm\omega} \right\}=\frac{2L}{\sigma_{\mathrm s}^2}\Re \left\{\frac{\partial \mathbf v_{\mathrm{s}}^\H}{\partial \bm\omega}\frac{\partial \mathbf v_{\mathrm{s}}}{\partial \bm\omega} \right\},
\end{equation}  
The detailed derivations for obtaining \eqref{eq:CRLB} from \eqref{FIM:original} are provided in Appendix \ref{App:derivation of FIM}.
\end{IEEEproof}

\subsection{Problem Formulation}
Define $\mathbf W=[\mathbf W_\rmc,\mathbf W_\rms]\in\mathbb C^{\Nt\times (K+\Ns)} $. Our goal is to optimize the transmit beamforming matrix $ \mathbf W $ to achieve a favorable tradeoff between communications and sensing performances. We evaluate communications performance using the SR, and we adopt $\mathrm{tr}(\mathbf{F}^{-1})$ as the sensing performance metric, following \cite{Li2008range}, as the diagonal entries of $\mathbf F^{-1}$ represent the CRLBs for the elements of $\bm\omega$. Several scalar mappings of these CRLBs have been explored in prior work \cite{chen2024transmitter,zhu2023integrated,Liu2024SNR, chen2024fast,wu2024global}. Among these approaches, minimizing $\mathrm{tr}(\mathbf{F}^{-1})$ is the more common choice for the CRLB metric \cite{Li2008range}.


The beamforming design problem can be formulated as:
 \begin{equation}\label{P1}
 \max_{ \mathbf W\in\mathcal S}\,\, \delta_\rmc\sum_{k=1}^K R_k-\delta_\rms \mathrm{tr}(\mathbf{F}^{-1}) 
 \end{equation}
where $\mathcal S\triangleq\{\mathbf W\in\mathbb C^{\Nt\times (K+\Ns)}\big|\mathrm{tr}(\mathbf W\mathbf W^\H)\leq P_{\mathrm{t}}\} $, with $P_{\mathrm{t}} $ representing the transmit power budget. Adjusting the values of the weights $ \delta_\rmc \geq 0$ and $\delta_\rms\geq 0$ enables a tradeoff between the communications and sensing performances. 

Problem \eqref{P1} is inherently NP-hard due to the presence of multiple non-convex fractional SINRs. Solving \eqref{P1} to find the global optimum would involve exponential computational complexity, as indicated in~\cite{wu2024global}. Moreover, the blockwise matrix variables in $\bf F$ present a significant challenge for optimization. In the face of these challenges, we employ the SCA framework to develop an efficient approach that ensures a locally optimal solution for \eqref{P1}, as detailed in the following section.


\section{ISAC Beamforming Optimization via SCA }
\label{sec: algorithm}
To deal with the non-convex objective function in \eqref{P1},  we develop a two-layer SCA approach, which allows a more tractable surrogate objective function and guarantees convergence. Before delving into the details,  we first present the full power consumption property in the following proposition.
%
\begin{proposition}\label{pro:full power}
    Any locally optimal point of problem \eqref{P1} must result in the full power consumption, i.e., satisfying $\mathrm{tr}(\mathbf W\mathbf W^\H) =  P_{\mathrm{t}}$. Thus, problem \eqref{P1} is equivalent to
    \begin{equation}\label{P1:boundary}
 \max_{ \mathbf W\in\mathcal B}\,\, \delta_\rmc\sum_{k=1}^K R_k-\delta_\rms \mathrm{tr}(\mathbf{F}^{-1}) 
 \end{equation}
 where $\mathcal B\triangleq\{\mathbf W\in\mathbb C^{\Nt\times (K+\Ns)}\big| \rmtr(\mathbf W\mathbf W^
\H)=P_{\mathrm t}\}$.
\end{proposition}
 \begin{IEEEproof}
   Please see Appendix \ref{property:full power}.
\end{IEEEproof}

\smallskip
The full power consumption property for the subproblem with $\delta_\rms=0$ has been presented in \cite{zhao2023rethink,fang2024rate}. In this work, we extend the property to ISAC systems. Intuitively, Proposition~\ref{pro:full power} suggests that higher transmit power leads to improved performance. As a result, the transmit beamforming matrix can be directly set to utilize the maximum power budget without incurring any performance loss. We will later employ the full power consumption property to facilitate our algorithm design.

\subsection{First-Layer SCA} \label{sec:SCA method}
We first recall two lemmas that form the foundation for approximating the non-convex objective function in  \eqref{P1}.
\begin{lemma}[\!\!\cite{fang2023optimal}]\label{lemma:lin app com}
    Function $\log\left( 1+\frac{|z|^2}{d}\right)$, $z\in\mathbb C, d\in\mathbb R_{++}$ is lower bounded by its first-order Taylor expansion as follows:
    \begin{equation}
        \begin{aligned}
            \log\left( 1+\frac{|z|^2}{d}\right )&\geq \log\left(1+\frac{|z_0|^2}{d_0}\right)+2\Re\{\frac{z_0^*}{d_0} z\}\\
            &-\frac{|z_0|^2}{d_0(d_0+|z_0|^2)}(|z|^2+d)-\frac{|z_0|^2}{d_0}.
        \end{aligned}
    \end{equation}
    Equality is achieved at $(z,d)=(z_0,d_0)$, where  $(z_0,d_0)$ is a given point that belongs to the domain. 
\end{lemma}

\begin{lemma}[\!\!\cite{sun2017major}] \label{lemma:lin app sen}
    The function $-\mathrm{tr}(\mathbf Z^{-1}), \mathbf Z\in \mathbb S_{++}$ is upper bounded by its first-order Taylor expansion as follows: 
    \begin{equation}
        -\mathrm{tr}(\mathbf Z^{-1})\leq \mathrm{tr}(\mathbf Z_0^{-1}\mathbf Z\mathbf Z_0^{-1} )-2\mathrm{tr}(\mathbf Z_0^{-1}).
    \end{equation}
    Equality is achieved at $\mathbf Z=\mathbf Z_0$, where $\mathbf Z_0$ is a given point that belongs to the domain.
\end{lemma}

Applying Lemma \ref{lemma:lin app com} to $R_k$ in \eqref{rate}, we construct a surrogate function at iteration $t$ as follows
\begin{align}\label{surrogate: com}
    r_k^{[t]}&\triangleq\log(1+\xi_{k}^{{[t]}})+2\Re\{\mathbf h_k^\H\mathbf w_{\rmc k}\eta_{ k}^{{[t]}} \} -\xi_{k}^{[t]}\nonumber \\
		&\quad -\beta_k^{{[t]}}\left(\sum_{j=1}^{K}|\mathbf h_k^\H\mathbf w_{\rmc j}^{[t]}|^2+\|\mathbf h_k^\H\mathbf W_\rms^{[t]}\|_\rmf^2+\sigma_{\mathrm{c} k}^2  \right),
\end{align}
where $\xi_k^{{[t]}}$, $\eta_k^{{[t]}}$, and $\beta_k^{{[t]}}$ are auxiliary variables given by
\begin{subequations}\label{update auxilary communications}
    \begin{align}
        \xi_k^{{[t]}}&\triangleq\frac{|\mathbf h_k^\H\mathbf w_{\rmc k}^{[t]}|^2}{\sum_{j=1,j\neq k}^K|\mathbf h_k^\H\mathbf w_{\rmc j}^{[t]}|^2+\|\mathbf h_k^\H\mathbf W_\rms^{[t]}\|_\rmf^2+\sigma_{\mathrm{c}k}^2},\\
        \eta_k^{{[t]}}&\triangleq\frac{\xi_k^{[t]}}{\mathbf h_k^\H\mathbf w_{\rmc k}^{[t]}},\\
        \beta_k^{{[t]}}&\triangleq\frac{\xi_k^{{[t]}}}{\sum_{j=1}^K|\mathbf h_k^\H\mathbf w_{\rmc j}^{[t]}|^2+\|\mathbf h_k^\H\mathbf W_\rms^{[t]}\|_\rmf^2+\sigma_{\mathrm c k}^2}.
    \end{align}
\end{subequations}
Similarly, by applying Lemma \ref{lemma:lin app sen} to $-\mathrm{tr}(\mathbf{F}^{-1})$, we construct the surrogate $\mathrm{tr}(\mathbf F \mathbf \Phi^{[t]})-2\mathrm{tr}(\mathbf F^{[t]})$ at iteration $t$,
where $\bm\Phi^{[t]}\triangleq{\mathbf F^{[t]}}^{-1}{\mathbf F^{[t]}}^{-1}$. With these surrogate functions and given a feasible point $\mathbf W^{[t]}$, problem \eqref{P1:boundary} can be expressed  as
 \begin{equation}
 	\label{P1:SCA}
 	\max_{ \mathbf W\in\mathcal B}\,\, \delta_\rmc\sum_{k=1}^K r_k^{[t]}+ \delta_\rms\left(\mathrm{tr}(\mathbf F\bm\Phi^{[t]})-2\mathrm{tr}(\mathbf F^{[t]})\right). 
 \end{equation}
 
Although \eqref{P1:SCA} is still non-convex, its objective function now has a quadratic form with respect to $\mathbf W$ on the sphere boundary $\mathcal B$. This problem can be handled efficiently via the SCA method as elaborated next. 

\begin{remark}
    We remark that the first term in \eqref{P1:SCA}, i.e., $\delta_\rmc\sum_{k=1}^K r_k^{[t]}$, serves as a lower bound for the original function $\delta_\rmc\sum_{k=1}^K R_k $, while the second term, i.e., $\mathrm{tr}(\mathbf F\bm\Phi^{[t]})-2\mathrm{tr}(\mathbf F^{[t]})$, acts as an upper bound for $-\rmtr(\mathbf F^{-1})$. However, their sum is not a lower bound of the original objective function. Thus, the proposed method differs from most existing SCA frameworks, which first construct a lower bound surrogate function for the original objective function and then solve the surrogate subproblem based on that lower bound \cite{sun2017major}. Nevertheless,
the convergence of the proposed method will be established in Section \ref{sec:convergence}.
\end{remark}


\subsection{Second-Layer SCA}
\label{sec: SGPI method}
The blockwise structure of $\bf F $ poses a significant challenge for the SCA-based optimization in \eqref{P1:SCA}. To address this issue, we first partition $\bm\Phi$ into blocks as
\begin{equation}
\bm\Phi=\begin{bmatrix} \mathbf\Phi_{11}  & \mathbf\Phi_{12}  &\mathbf\Phi_{13}  &\mathbf\Phi_{14}  \\ \mathbf\Phi_{12}^\T  & \mathbf\Phi_{22}  &\mathbf\Phi_{23}  &\mathbf\Phi_{24}  \\\mathbf\Phi_{13}^\T  & \mathbf\Phi_{23}^\T  &\mathbf\Phi_{33}  &\mathbf\Phi_{34}  \\\mathbf\Phi_{14}^\T  & \mathbf\Phi_{24}^\T  &\mathbf\Phi_{34}^\T  &\mathbf\Phi_{44}  \\ \end{bmatrix}, 
\end{equation}
which allows us to recast the objective function of \eqref{P1:SCA} in a more explicit and compact form, as stated in the following proposition.

\begin{proposition}\label{pro:compact_form}
    Problem \eqref{P1:SCA} can be recast as
    \begin{align}\label{P1SCA:compact form}
            \max_{\mathbf W\in\mathcal B}\ 
            &2\delta_\rmc\Re\{\rmtr(\mathbf W_\rmc\bm\Sigma_1^{[t]}\mathbf H^\H)\}+\delta_\rms\Re\{\rmtr(\mathbf W\mathbf W^\H\mathbf Q^{[t]})\} \nonumber\\
            &\qquad -\delta_\rmc \rmtr(\Rx\mathbf H\bm\Sigma_2^{[t]}\mathbf H^\H), 
    \end{align}
    where 
    \begin{align}
        \mathbf \Sigma_1^{[t]} &\triangleq\mathrm{diag}(\eta_1^{[t]},\ldots,\eta_K^{[t]}) \\ 
        \mathbf H &\triangleq[\mathbf h_1,\ldots,\mathbf h_K] \\
        \mathbf \Sigma_2^{[t]} &\triangleq\mathrm{diag}(\beta_1^{[t]},\ldots,\beta_K^{[t]}) \\
        \mathbf Q^{[t]} &\triangleq \frac{2L}{\sigma_s^2}(\mathbf Q_{11}^{[t]}+\mathbf Q_{12}^{[t]}+\mathbf Q_{13}^{[t]}+\mathbf Q_{22}^{[t]}+\mathbf Q_{23}^{[t]}+\mathbf Q_{33}^{[t]}) \; ,
     \end{align}
and $\mathbf Q_{11}^{[t]}, \mathbf Q_{12}^{[t]}, \mathbf Q_{13}^{[t]}, \mathbf Q_{22}^{[t]}, \mathbf Q_{23}^{[t]}, \mathbf Q_{33}^{[t]}$ are defined in \eqref{definition of q}. 
    
    \begin{figure*}[ht]
  \begin{align} \label{definition of q}
            &\mathbf Q_{11}^{[t]}\triangleq\mathbf A\mathbf U^\H(\mathbf \Phi_{11}^{[t]}\!\odot\!(\dot{\mathbf B}_{\bm\theta}^\H\dot{\mathbf B}_{\bm\theta})) \mathbf U\mathbf A^\H\!+\!\dot{\mathbf A}_{\bm\theta}\mathbf U^\H(\mathbf\Phi_{11}^{[t]}\!\odot\!({\mathbf B}^\H\dot{\mathbf B}_{\bm\theta}))  \mathbf U\mathbf A^\H \!+\!\mathbf A\mathbf U^\H(\mathbf\Phi_{11}^{[t]}\!\odot\!(\dot{\mathbf B}_{\bm\theta}^\H{\mathbf B}))   \mathbf U\dot{\mathbf A}_{\bm\theta}^\H\!+\!\dot{\mathbf A}_{\bm\theta}\mathbf U^\H  (\mathbf\Phi_{11}^{[t]}\!\odot\!({\mathbf B}^\H{\mathbf B}))    \mathbf U\dot{\mathbf A}_{\bm\theta}^\H\nonumber\\
&\mathbf Q_{12}^{[t]} \triangleq 2\big(\mathbf A\mathbf U^\H(\mathbf\Phi_{12}^{[t]}\!\odot\!(\dot{\mathbf B}_{\bm\theta}^\H\dot{\mathbf B}_{\bm\phi}) )\mathbf U\mathbf A^\H\!+\!\dot{\mathbf A}_{\bm\theta}\mathbf U^\H(\mathbf\Phi_{12}^{[t]}\!\odot\!({\mathbf B}^\H\dot{\mathbf B}_{\bm\phi}))\mathbf U\mathbf A^\H \!+\!\mathbf A\mathbf U^\H(\mathbf \Phi_{12}^{[t]}\!\odot\!(\dot{\mathbf B}_{\bm\theta}^\H{\mathbf B}))\mathbf U\dot{\mathbf A}_{\bm\phi}^\H\!+\!\dot{\mathbf A}_{\bm\theta}\mathbf U^\H(\mathbf \Phi_{12}^{[t]}\!\odot\!({\mathbf B}^\H{\mathbf B}))\mathbf U\dot{\mathbf A}_{\bm\phi}^\H\big) \nonumber\\
& \mathbf Q_{22}^{[t]} \triangleq\mathbf A\mathbf U^\H(\mathbf \Phi_{22}^{[t]}\!\odot\!(\dot{\mathbf B}_{\bm\phi}^\H\dot{\mathbf B}_{\bm\phi})) \mathbf U\mathbf A^\H\!+\!\dot{\mathbf A}_{\bm\phi}\mathbf U^\H(\mathbf\Phi_{22}^{[t]}\!\odot\!({\mathbf B}^\H\dot{\mathbf B}_{\bm\phi}))  \mathbf U\mathbf A^\H \!+\!\mathbf A\mathbf U^\H(\mathbf\Phi_{22}^{[t]}\!\odot\!(\dot{\mathbf B}_{\bm\phi}^\H{\mathbf B}))   \mathbf U\dot{\mathbf A}_{\bm\phi}^\H\!+\!\dot{\mathbf A}_{\bm\phi}\mathbf U^\H  (\mathbf\Phi_{22}^{[t]}\!\odot\!({\mathbf B}^\H{\mathbf B}))    \mathbf U\dot{\mathbf A}_{\bm\phi}^\H \nonumber\\
  &\mathbf Q_{13}^{[t]} \triangleq \mathbf A\mathbf U^\H((2\mathbf \Phi_{13}^{[t]}+2\mathrm j\mathbf\Phi_{14}^{[t]})\odot(\dot{\mathbf B}_{\bm\theta}^\H\mathbf B) )\mathbf A^\H +\dot{\mathbf A}_{\bm\theta}\mathbf U^\H((2\mathbf \Phi_{13}^{[t]}+2\mathrm j\mathbf\Phi_{14}^{[t]})\odot({\mathbf B}^\H\mathbf B))\mathbf A^\H  \nonumber \\
  &\mathbf Q_{23}^{[t]} \triangleq \mathbf A\mathbf U^\H((2\mathbf \Phi_{23}^{[t]}+2\mathrm j\mathbf\Phi_{24}^{[t]})\odot({\mathbf B}_{\bm\phi}^\H\mathbf B) )\mathbf A^\H +\dot{\mathbf A}_{\bm\phi}\mathbf U^\H((2\mathbf \Phi_{23}^{[t]}+2\mathrm j\mathbf\Phi_{24}^{[t]})\odot(\dot{\mathbf B}^\H\mathbf B))\mathbf A^\H  \nonumber \\
&\mathbf Q_{33}^{[t]}\triangleq\mathbf A((\mathbf\Phi_{33}^{[t]}+\mathbf\Phi_{44}^{[t]}+2\mathrm j\mathbf\Phi_{34}^{[t]})\odot(\mathbf B^\H\mathbf B))\mathbf A^\H .
\end{align}
    
\hrulefill 
\end{figure*}

\end{proposition}

 \begin{IEEEproof}
Please see Appendix \ref{App:quadratic problem}.    
\end{IEEEproof}

\smallskip
Problem \eqref{P1SCA:compact form} is a non-convex quadratically constrained quadratic problem (QCQP) with respect to $\mathbf W$. This stems from the fact that the matrix $\delta_\rms\mathbf Q^{[t]}-\delta_\rmc\mathbf H\mathbf \Sigma_2^{[t]}\mathbf H^\H$ is not necessarily negative definite and the feasible set $\mathcal B$ is not convex. To overcome this challenge and enable an efficient solution, we propose an equivalent but more tractable objective function. 

Specifically, considering that $\rm tr(\mathbf W\mathbf W^\H)= P_{\mathrm t}$ is satisfied at the optimal point, we can add $\lambda \left(\rm tr(\mathbf W\mathbf W^\H)- P_{\mathrm t}\right)$ to the objective function of \eqref{P1SCA:compact form}, where $\lambda$ is a predefined constant. With some algebra, problem \eqref{P1SCA:compact form} can be written as
      \begin{equation}\label{P1SCA:SQP}
           \max_{\mathbf W\in\mathcal B}\,\, 2\Re\{\rmtr(\mathbf W^\H \mathbf C_1^{[t]})\}+\rmtr(\Rx\mathbf C_{2}^{[t]}),
    \end{equation}
with $\mathbf C_1^{[t]}\triangleq[\delta_\rmc\mathbf H\mathbf \Sigma_1^{[t]^\H},\mathbf 0_{\Nt\times \Ns}]$ and  $  \mathbf C_2^{[t]}\triangleq\lambda\mathbf I+\frac{1}{2}\delta_\rms(\mathbf Q^{[t]}+\mathbf Q^{[t]^\H})-\delta_\rmc\mathbf H\mathbf\Sigma_2^{[t]}\mathbf H^\H$. 
By setting $\lambda$ as the absolute value of the dominant eigenvalue of $\delta_\rmc\mathbf H\mathbf\Sigma_2^{[t]}\mathbf H^\H -\frac{1}{2}\delta_\rms(\mathbf Q^{[t]}+\mathbf Q^{[t]^\H})$, then $\mathbf C_2^{[t]}$ is guaranteed to be a positive semidefinite matrix. With this condition, problem~\eqref{P1SCA:SQP} is in the form of the trust region problem (TRP) \cite{beck2018global} with the matrix variable $\mathbf W$. TRPs with a vector variable can be solved via methods such as bisection dual search \cite{fang2024BDRIS}, SDR \cite{rendl1997semi}, eigenvalue problem reformulation \cite{adachi2017solving}, and the first-order conic method \cite{beck2018global}. However, all of these approaches developed for vector-based TRPs require the eigenvalue decomposition, resulting in high computational complexity. To efficiently solve \eqref{P1SCA:SQP}, we recall the following Lemma.

\begin{lemma}[\!\!\cite{fang2024beamforming}]\label{lemma: trace quadratic lower bound}
For any given positive semidefinite Hermitian matrix $ \mathbf C \in\mathbb{C}^{N_{\mathrm{t}}\times N_{\mathrm{t}}}$ and any matrix $\mathbf W\in\mathbb C^{\Nt\times(K+\Ns)}$, the function $\mathrm{tr}(\mathbf W\mathbf W^\H\mathbf C)$ is lower bounded by its first-order Taylor expansion, given by 
\begin{equation}\label{linear lemma}
\mathrm{tr}(\mathbf W\mathbf W^\H\mathbf C)\geq 2\Re\{\mathrm{tr}(\mathbf W_0\mathbf W^\H\mathbf C) \}-\mathrm{tr}(\mathbf W_0\mathbf W_0^\H\mathbf C ),
\end{equation}
where $ \mathbf W_0$ is a given point that belongs to the domain, and equality is achieved if and only if $ \mathbf W=\mathbf W_0$.
\end{lemma}
With Lemma~\ref{lemma: trace quadratic lower bound}, we can further derive the following linear approximation for problem \eqref{P1SCA:SQP} at iteration $t$:
\begin{equation}\label{linear approximation}
	\max_{\mathbf W\in\mathcal B}\,\, 2\Re\{\mathrm{tr}(\mathbf W^\H\mathbf C_1^{[t]} +\mathbf W^\H\mathbf C_2^{[t]}\mathbf W^{[t]}) \}.
\end{equation}
Again, using the property $\rmtr(\mathbf W\mathbf W^\H)=P_{\mathrm t}$, problem \eqref{linear approximation} can be reformulated as the following projection problem
\begin{equation}\label{projection}
	\min_{\mathbf W\in\mathcal B}\,\, \|\mathbf W-(\mathbf C_1^{[t]}+\mathbf C_2^{[t]}\mathbf W^{[t]})\|_{\mathrm F}^2.
\end{equation}
The optimal solution to \eqref{projection} is known to be
\begin{equation}\label{Update W}
	\mathbf W^{[t+1]}= \bm\Pi_{\mathcal B}\left(\mathbf C_1^{[t]}+\mathbf C_2^{[t]} \mathbf W^{[t]} \right),
\end{equation}
where $ \bm\Pi_{\mathcal B}(\cdot) $ denotes the projection of its argument onto $ \mathcal B $, i.e., 
\begin{equation}
    \bm\Pi_{\mathcal B}({\mathbf X})\triangleq \sqrt{P_{\mathrm{t}}/\mathrm{tr}({\mathbf X} {\mathbf X}^\H)}{\mathbf X}.
\end{equation} 
\begin{remark}
    The proposed SCA method can be readily applied with per-antenna power constraints by replacing the total power constraint projection $\bm\Pi_{\mathcal B}(\cdot)$ with the per-antenna power constraint projection $\bm\Pi_{\mathcal P}(\cdot)$, given by
\begin{equation}
    \bm\Pi_{\mathcal P}(\mathbf X)\triangleq\sqrt{P_{\mathrm t}/\Nt}(\mathbf I\odot (\mathbf X\mathbf X^\H))^{-1} \mathbf X,
\end{equation}
where $\mathcal P\triangleq\{\mathbf W\in\mathbb C^{\Nt\times(K+\Ns)}\big| \mathrm{diag}(\Rx)= P_{\mathrm t} \mathbf 1_{\Nt}/\Nt\} $. Here, $\mathrm{diag}(\bf Y)$ returns a vector whose elements are the diagonal entries of $\bf Y$
\end{remark}

\subsection{Overall Optimization Framework for Problem \eqref{P1} }

Combining the proposed first- and second-layer SCA method in Sections \ref{sec:SCA method} and \ref{sec: SGPI method} respectively, we summarize the overall optimization algorithm for solving problem \eqref{P1} in Algorithm \ref{alg: SCA-SGPI}. Starting with an initial non-zero feasible point $\mathbf W^{[0]}$, the auxiliary variables $\xi_k,\eta_k,\beta_k,\forall k\in\mathcal K$ and $\bm\Phi$ in the first-layer SCA framework are updated based on \eqref{update auxilary communications} and \eqref{eq:CRLB}. Then, the auxiliary variables $\mathbf C_1$ and $\mathbf C_2$ are updated based on their definitions in \eqref{P1SCA:SQP}. Finally, $\mathbf W$ is obtained with \eqref{Update W}. This process continues until the objective value in \eqref{P1} converges.

\begin{algorithm}
	\textbf{Initialize}: $t\leftarrow0$, $\mathbf{W}^{[0]}$\;
	\Repeat{The objective value in \eqref{P1} converges}{
		$t\leftarrow t+1$\;
		Update $\xi_k^{[t]}, \eta_k^{[t]} $ and $ \beta_k^{[t]}$ according to \eqref{update auxilary communications} for $\forall k\in\mathcal K$\;
		Update $\mathbf \Phi^{[t]}={\mathbf F^{[t]}}^{-2}$ according to \eqref{eq:CRLB}\;	
	
 Update $\mathbf W^{[t]} $ according to \eqref{Update W}\;
 	
	}	
	\caption{The proposed SCA Algorithm}
	\label{alg: SCA-SGPI}				
\end{algorithm}

\subsection{Convergence Analysis}
\label{sec:convergence}

In this subsection, we show that the objective value monotonically increases after each iteration of Algorithm \ref{alg: SCA-SGPI}, leading to  convergence. The core idea of this convergence analysis is to map the proposed SCA algorithm into a projected gradient ascent (PGA) algorithm \cite{davis2019stochastic}. Specifically, we denote the original objective function in problem \eqref{P1:boundary} as $f(\mathbf W)\triangleq f_\rmc(\mathbf W)+f_\rms(\mathbf W)$ with $f_\rmc(\mathbf W)\triangleq\delta_\rmc\sum_{k=1}^K R_k$ and $f_\rms(\mathbf W)\triangleq-\delta_\rms\rmtr(\mathbf F^{-1})$, respectively.
Similarly, the objective function in \eqref{linear approximation} can be expressed as $h^{[t]}(\mathbf W)\triangleq h^{[t]}_\rmc(\mathbf W)+h_\rms^{[t]}(\mathbf W)+\lambda \rmtr(\mathbf W^\H\mathbf W^{[t]})$ with $h^{[t]}_\rmc(\mathbf W)=\Re\{\rmtr(2\mathbf W^\H\mathbf C_1^{[t]}-\delta_\rmc\mathbf W^\H\mathbf H\mathbf\Sigma_2^{[t]}\mathbf H^\H\mathbf W^{[t]})\}$ and $h_\rms^{[t]}(\mathbf W)=\frac{1}{2}\delta_\rms\rmtr(\mathbf W^\H(\mathbf Q^{[t]}+\mathbf Q^{[t]^\H})\mathbf W^{[t]})$. Then we have the following proposition.

\begin{proposition}\label{pro:gradient}
    The gradient of $f(\mathbf W)$ with respect to $\mathbf W$ at $\mathbf{W}=\mathbf{W}^{[t]}$ can be expressed as
    \begin{align}
            & \nabla f(\mathbf{W})\big|_{\mathbf{W}=\mathbf{W}^{[t]}} \notag \\
             &\quad =\nabla h_\rmc^{[t]}(\mathbf W)\big|_{\mathbf{W}=\mathbf{W}^{[t]}}+\nabla h_\rms^{[t]}(\mathbf W)\big|_{\mathbf{W}=\mathbf{W}^{[t]}} \notag \\
             &\quad =2\mathbf C_1^{[t]}+(\delta_\rms\mathbf Q^{[t]}+\delta_\rms\mathbf Q^{[t]^\H}-2\delta_\rmc \mathbf H\mathbf \Sigma_2^{[t]}\mathbf H^\H)\mathbf W^{[t]}.\label{gradient of f} 
    \end{align}
\end{proposition}

\begin{IEEEproof}
Please see Appendix \ref{app:gradient}.
\end{IEEEproof}

\smallskip
With Proposition \ref{pro:gradient}, \eqref{Update W} can be rewritten as
\begin{align} 
        \mathbf W^{[t+1]}&=\mathbf\Pi_{\mathcal B}\bigg(\mathbf C_1^{[t]}+\lambda\mathbf W^{[t]}+\frac{\delta_{\rm s}}{2}\big(\mathbf Q^{[t]}+\mathbf Q^{[t]^\H}\big)\mathbf W^{[t]}\notag \\
        &\quad\quad\quad\quad-\delta_{\rm c} \mathbf H\mathbf \Sigma_2^{[t]}\mathbf H^\H\mathbf W^{[t]}\bigg)\notag \\
        &=\mathbf\Pi_{\mathcal B}\left(\mathbf W^{[t]}+\frac{1}{2\lambda}\nabla f(\mathbf{W})\big|_{\mathbf{W}=\mathbf{W}^{[t]}} \right), \label{projection update}
\end{align}
where we have utilized the property $\mathbf\Pi_{\mathcal B}(\lambda \mathbf X)=\mathbf\Pi_{\mathcal B}( \mathbf X)$
for any constant $\lambda>0$.
Therefore, Algorithm \ref{alg: SCA-SGPI} can be interpreted as a PGA algorithm with step size $\frac{1}{2\lambda}$. In this context, constructing the surrogate functions in the first- and second-layer SCA  corresponds to computing the gradient of the original objective function $f(\mathbf W)$, while choosing an appropriate value of $\lambda$ is analogous to determining a suitable step size. With these observations, we establish the convergence of Algorithm \ref{alg: SCA-SGPI} as elaborated below. 
\begin{proposition}\label{pro:convergence}
    Assume that the objective function $f(\mathbf W)$ is $\rho$-weakly convex \cite{chen2022distributed}. When the shift parameter satisfies $\lambda>\frac{\rho}{2}$, Algorithm \ref{alg: SCA-SGPI} generates a convergent sequence $\{f(\mathbf W^{[t]})\}$. 
\end{proposition}

\begin{IEEEproof}
Please  see Appendix \ref{APP:convergence}.
\end{IEEEproof}

\smallskip
 We note that although Proposition~\ref{pro:convergence} is developed for Algorithm~\ref{alg: SCA-SGPI}, it can be generalized to tackle the optimization problem with PGA,  where both the objective function and feasible set are non-convex. While Proposition~\ref{pro:convergence} introduces an additional assumption, requiring $f(\mathbf W)$ to be $\rho$-weakly convex, this requirement is quite reasonable in ISAC systems, as detailed in the following remark.
 
\begin{remark}
     Assume the channel gain is finite for each user, i.e., $\|\mathbf h_k\|<\infty, \forall k\in\mathcal K$, the radar cross-section $\alpha_m $ is positive and finite, i.e., $0<\alpha_m<\infty, \forall m\in\mathcal M$, and the noise power satisfies $\sigma_{\rmc,k}^2>0,\forall k\in\mathcal K$ and $ \sigma_{\rms}^2>0$. Since $\mathcal B$ is closed and compact, the gradient $\nabla f(\mathbf W)$ is bounded for any point $\mathbf W\in\mathcal B$. Thus there exists a constant $C>0$ such that $\|\nabla f(\mathbf Z)-\nabla f(\mathbf W)\|_\rmf\leq C\|\mathbf Z-\mathbf W\|_\rmf$, for any $\mathbf Z,\mathbf W\in\mathcal B$. This implies that $\nabla f(\mathbf W)$ is a $C$-smooth function, which satisfies the following inequality \cite{nesterov2018lectures}
    \begin{equation}\label{C-smooth bound}
        f(\mathbf Z)\geq f(\mathbf W)+\langle \nabla f(\mathbf W), \mathbf Z-\mathbf W\rangle -\frac{C}{2}\|\mathbf Z-\mathbf W\|_\rmf^2,
    \end{equation}
    where $\langle\cdot,\cdot\rangle$ denotes the inner product.
    Since \eqref{C-smooth bound} is both a sufficient and necessary condition for $f(\mathbf W)$ to be weakly convex, we can conclude that $f(\mathbf W)$ is a $\rho$-weakly convex function with $\rho=C$ \cite[Lemma 1]{NEURIPS2019_05d0abb9}. While it is generally challenging to analytically determinate the specific parameter $\rho$, it can be estimated numerically by computing the dominant eigenvalue of the matrix $\mathbf C^{[t]}_2$ in each iteration. 
\end{remark}

\subsection{Computational Complexity Analysis}
The complexity of Algorithm \ref{alg: SCA-SGPI} is mainly due to the matrix multiplications and inverses. In each iteration, the complexity of calculating the auxiliary variables $\xi_k^{[t]}, \eta_k^{[t]}$, and $\beta_k^{[t]}$ is $\mathcal O(K^2\Nt+K\Nt^2)$. In step 5, the computation of the auxiliary variable $\bm\Phi^{[t]}$ requires a complexity of $\mathcal O(M^3+M^2\Nr+M^2\Nt+\Ns\Nt^2) $. The complexity of updating the beamforming matrix is $\mathcal O(M^2\Nr+M^2\Nt+K\Nt^2+\Ns\Nt^2)$. Consequently, the overall complexity of $I$ iterations of Algorithm \ref{alg: SCA-SGPI} is of the order of $\mathcal O\big(I\cdot(M^3+M^2(\Nt+\Nr)+K^2\Nt+K\Nt^2+\Ns\Nt^2)\big)$ operations.

\section{Optimal beamforming Structure}\label{Sec: OBS}

While Section~\ref{sec: algorithm} develops a SCA-based numerical algorithm for beamforming design, the structural properties of both the problem and its solution remain unclear, limiting deeper insights. Prior studies on the OBS in communications-only systems have demonstrated its advantages, including providing fundamental insights into beamforming directions \cite{emil2014optimal}, offering theoretical guarantees for optimal structures in extreme scenarios \cite{emil2014optimal}, enabling the design of low-complexity beamforming algorithms \cite{zhao2023rethink, fang2024rate}, and serving as a foundation for deep unfolding methods \cite{wang2024RSBNN,xia2020deep}. Motivated by these benefits, we seek to identify and exploit the OBS for problem \eqref{P1}. To that aim, we first derive the OBS and then establish a low-dimensional beamforming optimization framework.



\subsection{OBS for Problem~\eqref{P1}}
The Lagrangian function for problem $\eqref{P1}$ is expressed as
\begin{equation}\label{Lagrangian}
    \mathcal L(\mathbf W,\mu)=\delta_\rmc\sum_{k=1}^K R_k-\delta_\rms\rmtr(\mathbf F^{-1})-\mu(\rmtr(\mathbf W\mathbf W^\H)-P_{\mathrm t}),
\end{equation}
where $\mu$ is the Lagrange multiplier associated with the power constraint. The first-order optimal condition implies that for any stationary point $\mathbf W^\diamond$, there exists a corresponding Lagrange multiplier $\mu^\diamond$ that satisfies the first-order stationary condition, i.e., ${\partial \mathcal L}(\mathbf W,\mu)/{\partial \mathbf W}=\mathbf 0 $, which leads to
\begin{equation}\label{gradient equal zero}
 2 \mathbf C_1^\diamond+(\delta_s\mathbf Q^\diamond+\delta_s\mathbf Q^{\diamond^\H}-2\delta_c \mathbf H\mathbf \Sigma_2^\diamond\mathbf H^\H)\mathbf W^\diamond-2\mu^\diamond\mathbf W^\diamond=\mathbf 0,
\end{equation}
where the superscript $(\cdot)^\diamond$ highlights the values of the respective variables at the stationary point of problem \eqref{P1}. Note that $\mathbf C_1^\diamond$, $\mathbf Q^{\diamond} $, and $ \bm\Sigma_2^{\diamond}$ are functions of $\mathbf W^\diamond$ and their exact numerical values are not considered here since we are investigating the structure of the solutions.

By splitting $\mathbf W^\diamond$ into $\mathbf W_\rmc^\diamond$ and $\mathbf W_\rms^\diamond$, we obtain the optimal beamforming structure for CRLB optimization in ISAC systems, as characterized by the following Theorem \ref{OBS}. 
\begin{theorem}
\label{OBS}
    The optimal beamforming structure for the ISAC system with the given CRLB metric is \begin{align}\label{eq:OBS}
          &\mathbf W_\rmc^\diamond =\left(\mu^\diamond \mathbf I+\delta_\rmc\mathbf H\mathbf \Sigma_2^\diamond\mathbf H^\H-\frac{1}{2}\delta_\rms(\mathbf Q^\diamond+\mathbf Q^{\diamond^\H})\right)^{-1}\!\!\!\delta_\rmc\mathbf H\mathbf \Sigma_1^{\diamond^\H},\nonumber\\
       &\left(\frac{1}{2}\delta_\rms(\mathbf Q^\diamond+\mathbf Q^{\diamond^\H})-\delta_\rmc\mathbf H\mathbf \Sigma_2^\diamond\mathbf H^\H\right) \mathbf W_\rms^\diamond =\mu^\diamond \mathbf W_\rms^\diamond.
    \end{align}
\end{theorem}

\begin{IEEEproof}
    By dividing $\mathbf W^\diamond$ into $\mathbf W_\rmc^\diamond$ and $\mathbf W_\rms^\diamond$, we obtain  the following set of equations based on \eqref{gradient equal zero}:
    \begin{align}
          &2\delta_c\mathbf H\mathbf\Sigma_1^{\diamond^\H}\!+\delta_s(\mathbf Q^\diamond\!+\mathbf Q^{\diamond^\H}\!-2\delta_c \mathbf H\mathbf \Sigma_2^\diamond\mathbf H^\H)\mathbf W_\rmc^\diamond\!-2\mu^\diamond\mathbf W_\rmc^\diamond\!=\mathbf 0,\nonumber \\
       &\left(\delta_\rms(\mathbf Q^\diamond+\mathbf Q^{\diamond^\H})-2\delta_\rmc\mathbf H\mathbf \Sigma_2^\diamond\mathbf H^\H\right) \mathbf W_\rms^\diamond -2\mu^\diamond \mathbf W_\rms^\diamond=\mathbf 0,\nonumber
    \end{align}
   from which Theorem \ref{OBS} directly follows.
\end{IEEEproof}

\smallskip
It can be observed that the communications beamforming matrix $\mathbf W_\rmc$ generalizes the existing downlink multiuser unicast beamforming structure presented in \cite{emil2014optimal} to ISAC systems. Moreover, the columns of the optimal sensing beamforming matrix are either the eigenvectors of the matrix $ \frac{1}{2}\delta_\rms(\mathbf Q+\mathbf Q^\H)-\delta_\rmc\mathbf H\mathbf \Sigma_2\mathbf H^\H$ or the zero vector.

\subsection{Inherent Low-Dimensional Beamforming Structure}
Theorem~\ref{OBS} indicates that the optimal beamforming solutions can be fully determined by a set of parameters with significantly reduced dimensions. Specifically, instead of optimizing the original variable $\mathbf{W}$, which comprises $\Nt \times (K + \Ns)$ complex entries, it suffices to identify one real dual variable $\mu^\diamond$, $2K$ complex auxiliary variables in $\mathbf{\Sigma}_1^\diamond$ and $\mathbf{\Sigma}_2^\diamond$, and $8M^2$ real auxiliary variables in the symmetric matrix $\mathbf{\Phi}^\diamond$ on which $\mathbf{Q}^\diamond$ depends. This highlights that the dimension of the auxiliary  variables $\{\mu^\diamond,\mathbf{\Sigma}_1^\diamond,\mathbf{\Sigma}_2^\diamond,\mathbf{\Phi}^\diamond\}$ for obtaining the optimal beamforming solution $\{\mathbf W_\rmc^\diamond,\mathbf W_\rms^\diamond\}$ does not scale with the number of transmit antennas $\Nt$ or the number of sensing streams $\Ns$. 


Inspired by this fact, we now explore the inherent low-dimensional structure of \eqref{eq:OBS}. First, we provide an upper bound on the number of sensing streams $\Ns$ in the following lemma.
\begin{lemma}\label{pro:rank ws}
    With $\delta_c=0$, problem \eqref{P1} reduces to a CRLB minimization problem. The rank of its optimal solution is not more than $3M$, i.e., $\mathrm{rank}(\mathbf W_s^\diamond\big|_{\delta_c=0})\leq 3M$.
\end{lemma}
\begin{IEEEproof}
Please see Appendix \ref{app:rank ws}.
\end{IEEEproof}

\smallskip
Lemma \ref{pro:rank ws} indicates that $\Ns=3M$ is sufficient to fully exploit the available DoFs for sensing, and setting $\Ns>3M$ is unnecessary. Our subsequent simulation results verify this in Section \ref{Sec:number of streams}.

\subsection{Low-Dimensional Beamforming Design}

The above finding reveals that the dimensions of the optimization variable can be further reduced without compromising the performance. This motivates the following proposition.
\begin{proposition}\label{pro:ld structure}
    Problem \eqref{P1} can be reformulated as
     \begin{subequations}\label{P1:ld}
\begin{align}
    \max_{ \mathbf P}\,\, &\delta_\rmc\sum_{k=1}^K R_k-\delta_\rms \mathrm{tr}(\mathbf{F}^{-1}) \\
    \mathrm{s.t.}\,\, 
    & \mathbf W=\mathbf N\mathbf P,\\
    &\rmtr(\mathbf P\mathbf P^\H\mathbf N^\H\mathbf N)\leq P_\mathrm{t},
\end{align} 
 \end{subequations}
 where $\mathbf P=[\mathbf P_\rmc,\mathbf P_\rms]\in\mathbb C^{(K+3M)\times (K+3M)}$ is an auxiliary variable and $\mathbf N=[\mathbf H, \mathbf A, \dot{\mathbf A}_\theta, \dot{\mathbf A}_\phi]$.
\end{proposition}
\begin{IEEEproof}
Please see Appendix \ref{app:ld structure}.
\end{IEEEproof}


\smallskip
The number of variables in \eqref{P1:ld} is $K^2+9M^2+6MK$, which is significantly lower than that of the original problem \eqref{P1} when $\Nt\gg \max\{K,M\}$. Furthermore, problem \eqref{P1:ld} has a structure similar to the original problem \eqref{P1}, enabling an efficient low-complexity solution. Specifically, by defining the effective channel matrix as $[\widetilde{\mathbf H},\widetilde{\mathbf A}, \widetilde{\dot{\mathbf A}}_\theta,\widetilde{\dot{\mathbf A}}_\phi]  =\mathbf N^\H[\mathbf H, \mathbf A, \dot{\mathbf A}_\theta, \dot{\mathbf A}_\phi]$ and substituting the original channel matrices $\{\mathbf H, \mathbf A, \dot{\mathbf A}_\theta, \dot{\mathbf A}_\phi\}$ with their transformed counterparts $\{\widetilde{\mathbf H},\widetilde{\mathbf A}, \widetilde{\dot{\mathbf A}}_\theta,\widetilde{\dot{\mathbf A}}_\phi\}$ in problem \eqref{P1:ld}, we obtain a reformulated problem that retains the structure of the original one. 

Consequently, the methods used to solve for the original beamforming matrix $\mathbf W$ can be similarly applied in computing $\mathbf P$. The key difference is that after the first- and second-layer SCA, the following linear approximation problem is obtained at iteration~$t$
\begin{equation}\label{Plinear:ld}
\begin{split}
     \max_{\mathbf P}\,\,&\Re\left\{\rmtr\left(\mathbf P^\H\left(\widetilde{\mathbf C}_1^{[t]}+\widetilde{\mathbf C}_2^{[t]}\mathbf P^{[t]}\right)\right)\right\}\\\text{s.t.}\,\,&\rmtr(\mathbf P\mathbf P^\H\mathbf N^\H\mathbf N)=P_{\rm t},
\end{split}
\end{equation}
where $\widetilde{\mathbf C}_1^{[t]}$ and $\widetilde{\mathbf C}_2^{[t]}$ are auxiliary variables computed in the same way as $\mathbf C_1$ and $\mathbf C_2$ according to $\{\widetilde{\mathbf H},\widetilde{\mathbf A}, \widetilde{\dot{\mathbf A}}_\theta,\widetilde{\dot{\mathbf A}}_\phi\}$. Applying the Lagrangian method, we obtain a closed-form solution to \eqref{Plinear:ld}, written as
\begin{equation}
        \mathbf P^{[t+1]}=\sqrt{{P_{\mathrm t}}/{\rmtr(\mathbf L^{[t]}\mathbf L^{[t]^\H}(\mathbf N^{\H}\mathbf N)^{-1})}}(\mathbf N^{\H}\mathbf N)^{-1}\mathbf L^{[t]},
\end{equation}
with $\mathbf L^{[t]}=\widetilde{\mathbf C}_1^{[t]}+\widetilde{\mathbf C}_2^{[t]}\mathbf P^{[t]}$. We refer to the algorithm for solving the low-dimensional problem \eqref{P1:ld} as the low-dimensional Algorithm \ref{alg: SCA-SGPI}. This low-dimensional reformulation is particularly beneficial in extremely large MIMO systems \cite{zhang2022fast}.

\section{Simulation Results}\label{simulation}\label{Sec:simulation}
 In this section, we evaluate the computational complexity, convergence, and performance of the proposed algorithms.\footnote{The source code is available online at \url{https://github.com/Nostalgia2022/OBS-for-CRLB-ISAC}.} In all simulations, unless otherwise stated, we set $\Nt=16$, $\Nr=20$, $\Ns=\Nt$, $K=4$, $M=2$, $L=64$, $\delta_\rms=1$, and $\delta_\rmc=0.25$ \cite{liu2021cramer}. The transmit power is set to $ P_{\mathrm t}=10$ dBm while the noise variances are $\sigma_\rms^2=\sigma_{\rmc k}^2=0$ dBm. We adopt a Rayleigh fading model for the communications channel \cite{liu2021cramer}. The UPA for downlink communications has a size of $N_{\mathrm t\mathrm h}\times N_{\mathrm{tv}}$, where $N_{\mathrm t\mathrm h}=4$ and $N_{\mathrm{tv}}=4$ are the numbers of antennas in the horizontal and vertical dimensions, respectively.  The UPA steering vector $\mathbf a(\theta,\phi)$ is modeled as \cite{nhan2022UPA}  
 \begin{equation*}
     \mathbf a(\theta,\phi)=\mathbf a_{\mathrm h}(\theta,\phi)\otimes \mathbf a_{\mathrm v}(\phi),
 \end{equation*} 
with
\begin{equation*}
    \begin{aligned}
        &\mathbf a_{\mathrm h}(\theta,\phi)=\frac{1}{\sqrt{N_{\mathrm t\mathrm h}}}\left[1,e^{j\pi\sin{\theta}\sin{\phi}},\ldots,e^{j\pi(N_{\mathrm t\mathrm h}-1)\sin{\theta}\sin{\phi}}\right]^\T,\\
       & \mathbf a_{\mathrm v}(\phi)=\frac{1}{\sqrt{N_{\mathrm{tv}}}}\left[ 1,e^{j\pi\cos{\phi}},\ldots,e^{j\pi(N_{\mathrm t\mathrm v}-1)\cos{\phi}}\right]^\T,
    \end{aligned}
\end{equation*}
where $\otimes $ represents the Kronecker product. The partial derivatives of the steering vector $\mathbf a(\theta,\phi)$ with respect to $\theta$ and $\phi$ are expressed as
\begin{equation*}
    \begin{aligned}
       & \frac{\partial \mathbf a(\theta,\phi)}{\partial \theta}=\frac{\partial \mathbf a_{\mathrm h}(\theta,\phi)}{\partial \theta} \otimes \mathbf a_{\mathrm v}(\phi),\\
        & \frac{\partial \mathbf a(\theta,\phi)}{\partial \phi}=\frac{\partial \mathbf a_{\mathrm h}(\theta,\phi)}{\partial \phi} \otimes \mathbf a_{\mathrm v}(\phi)+\mathbf a_{\mathrm h}(\theta,\phi)\otimes \frac{\partial \mathbf a_{\mathrm v}(\phi)}{\partial \phi},
    \end{aligned}
\end{equation*}
where 
\begin{equation*}
    \begin{aligned}
       & \frac{\partial \mathbf a_{\mathrm h}(\theta,\phi)}{\partial \theta}=j\pi\cos{\theta}\sin{\phi}\bf n_{\rm th} \odot \mathbf a_{\mathrm h}(\theta,\phi),\\
       & \frac{\partial \mathbf a_{\mathrm h}(\theta,\phi)}{\partial \phi}=j\pi \sin{\theta}\cos{\phi}\bf n_{\rm th}\odot \mathbf a_{\mathrm h}(\theta,\phi),\\
       &\frac{\partial \mathbf a_{\mathrm v}(\phi)}{\partial \phi}=-j\pi\sin{\phi}\bf n_{\rm tv}\odot \mathbf a_{\mathrm v}(\phi),
    \end{aligned}
\end{equation*}
with ${\bf n}_{\rm th}=[0,1,\ldots,N_{\mathrm t\mathrm h}-1]^\T$ and ${\bf n}_{\rm tv}=[0,1,\ldots,N_{\mathrm t\mathrm v}-1]^\T$.
The steering vector $\mathbf b(\theta,\phi)$ and its partial derivatives are modeled similarly. Furthermore, the radar cross-section $\alpha_m$ is set as $\alpha_m=0.1\times(1+0.2\nu_{u})e^{2\mathrm j \pi \nu_{u}}$ \cite{fang2024beamforming}, where $\nu_u$ follows a uniform distribution $ \mathcal {U}(0,1)$. The angles $\theta_m$ and $\phi_m$ are sampled independently from a uniform distribution $\mathcal U(-2\pi/3,2\pi/3)$. The convergence tolerance for Algorithm \ref{alg: SCA-SGPI} is set to $10^{-4}$, and all the presented results are averaged over 100 channel realizations.

\begin{figure}[t]
\center
\includegraphics[width=0.45\textwidth]{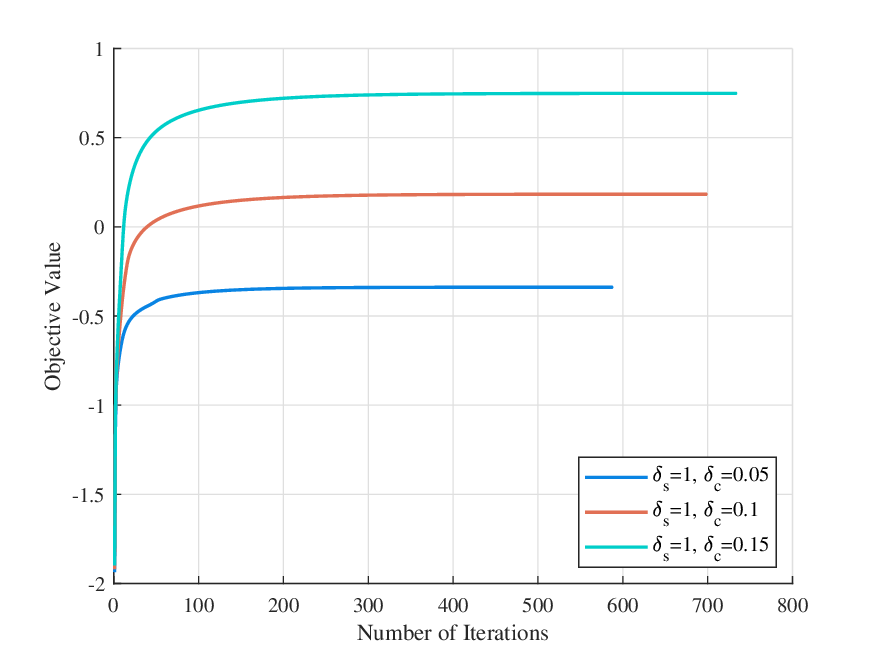}
    \vspace{-2mm}
    \caption{Convergence behavior of the proposed algorithm with various weight coefficients.}
    \label{fig:convergence}
\end{figure}
\begin{figure}[t]
\center
\includegraphics[width=0.45\textwidth]{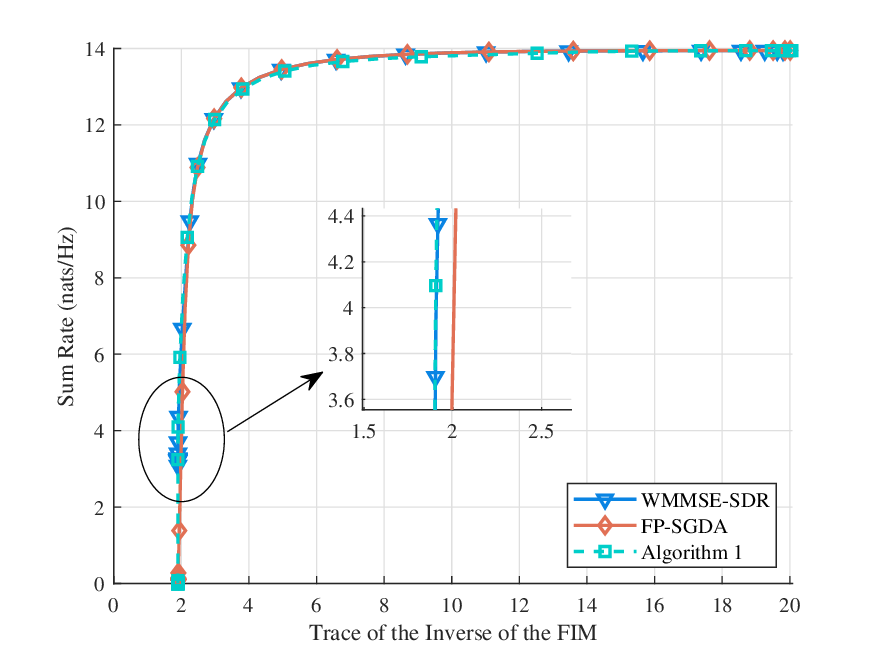}
    \vspace{-2mm}
    \caption{Tradeoff region between communications and sensing.}
    \label{fig:tradeoff}
\end{figure}

\subsection{Convergence and Performance}
We first evaluate the convergence behavior of Algorithm \ref{alg: SCA-SGPI} with different values of the weights $\delta_{\rm s}$ and $\delta_{\rm c}$, as shown in Fig.~\ref{fig:convergence}. We see that Algorithm \ref{alg: SCA-SGPI} exhibits a monotonic increase until convergence for all weight coefficients, which is consistent with the theoretical analysis presented in Proposition~\ref{pro:convergence}. Although Algorithm \ref{alg: SCA-SGPI} requires hundreds of iterations to converge, its per-iteration complexity is very low, leading to fast convergence in terms of CPU time.

In Fig.\ \ref{fig:tradeoff}, we show the communications–sensing tradeoff region achieved by the proposed approach compared against the WMMSE-SDR and FP-SGDA baselines from \cite{zhu2023integrated}. The WMMSE-SDR method ensures a locally optimal solution at the expense of a high computational complexity, while FP-SGDA offers a lower-complexity alternative with a slight performance loss. The tradeoff regions are generated by fixing $\delta_s = 1$ and varying $\delta_c$ from $10^{-7}$ to $10^5$. Fig.\ \ref{fig:tradeoff} shows that the proposed algorithm achieves a tradeoff region comparable to WMMSE-SDR. However, due to inherent differences in methodology, the two algorithms do not necessarily reach the same point on the tradeoff region boundary for identical weight coefficients. 

\subsection{Impact of Number of Radar Signal Streams $(\Ns)$}\label{Sec:number of streams}
\begin{figure}[t]
\center
\includegraphics[width=0.45\textwidth]{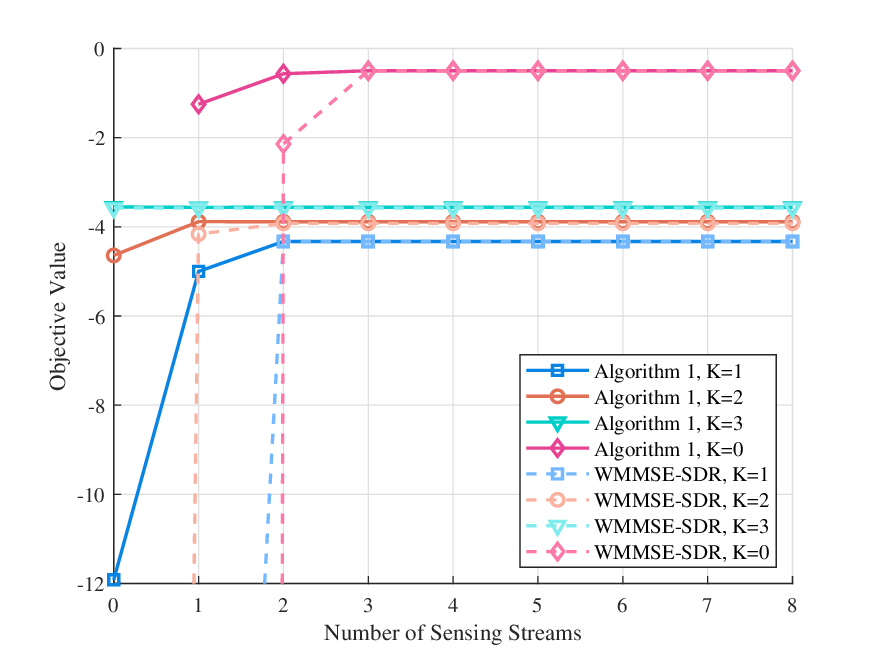}
    \vspace{-2mm}
    \caption{Objective value versus the number of sensing streams.}
    \label{fig:Ns}
\end{figure}
 Fig.\ \ref{fig:Ns} illustrates the value of Algorithm \ref{alg: SCA-SGPI}'s objective function as $\Ns$ increases, with $L=128$ and $M=3$. The results show that the objective increases until $\Ns$ reaches a specific threshold. For a sensing-only system, the threshold is $\Ns=M=3$, which is significantly lower than the theoretical upper bound $3M$ presented in Lemma \ref{pro:rank ws}. For ISAC systems, the threshold decreases as the number of communications users increases. Notably, when $\Ns=\max\{0, M-K\}$, the performance ceases to improve with further increases of $\Ns$. This observation reveals that only $M-K$ data streams (rather than $\Nt$ as used in \cite{liu2020joint}) are needed to fully exploit the sensing DoFs. In particular, when $K \geq M$, i.e., the number of users is not less than the number of targets, it is sufficient to use only information signals $ \ssb_{\mathrm c}[l],\forall l$, to guarantee the sensing performance. Furthermore, this highlights the flexibility of Algorithm \ref{alg: SCA-SGPI}, as it consistently achieves the same or better performance across all values of $\Ns$. Note that the beamforming matrix $\mathbf W_\rms$ has dimension $\Nt\times \Ns$, whereas the SDR-based algorithms lift it to the covariance domain, where $\mathbf R_\rms=\mathbf W_\rms\mathbf W_\rms^\H$ has dimension $\Nt\times \Nt$ regardless of $\Ns$. This fundamental difference makes our algorithm significantly more efficient than the SDR-based approaches. We can also observe that WMMSE-SDR incurs a significant performance loss when the number of sensing streams is insufficient.


\subsection{Comparison With Existing Locally Optimal Algorithms}

\begin{figure}[t]
\center
\includegraphics[width=0.45\textwidth]{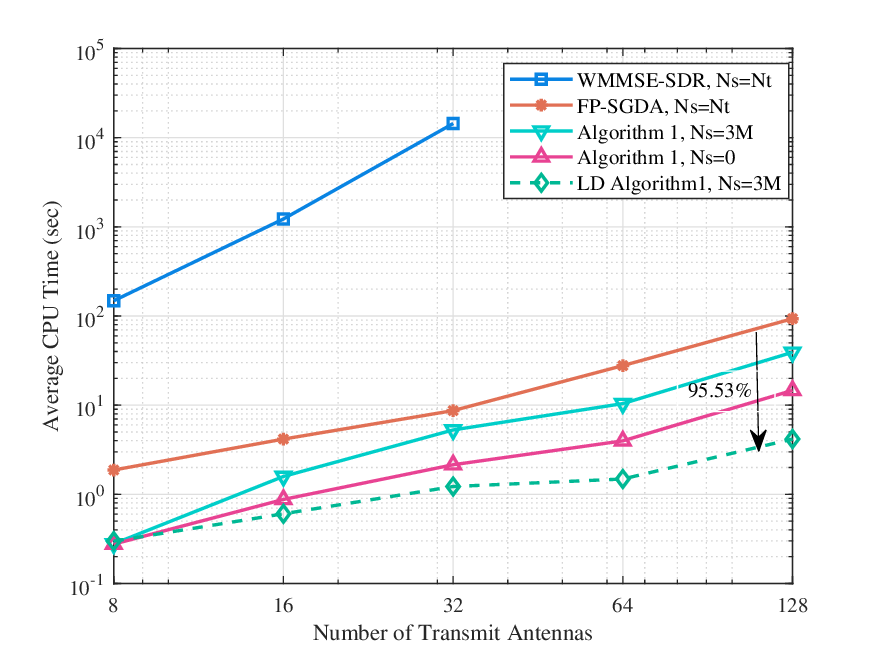}
    \vspace{-2mm}
    \caption{Average CPU Time vs. the number of transmit antennas.}
    \label{fig:Nt}
\end{figure}
\begin{table*}[t]
\small
    \centering
    \caption{Sum Rate and CRLB Performance comparison for different beamforming algorithms.}
    \label{tab:comparison}
    \begin{tabular}{lccccc}
        \toprule
        \multirow{2}{*}{Algorithm} & \multicolumn{5}{c}{Number of Transmit Antennas $N_t$} \\
        \cmidrule(lr){2-6}
         & $\Nt = 8$ & $\Nt = 16$ & $\Nt = 32$ & $\Nt = 64$ & $\Nt=128$ \\
        \midrule
        \multicolumn{6}{l}{\textit{Each entry represents (Sum Rate, CRLB).}} \\
        WMMSE-SDR   & (11.02, 2.50)  & (15.02, 1.12) & (18.16, 0.83)&N/A & N/A\\
        FP-SGDA &(10.88, 2.47)&(14.97, 1.15)&(18.12, 0.87)&(20.91, 0.79)& (23.45, 0.82)\\
        Algorithm \ref{alg: SCA-SGPI}, $\Ns=3M$     & (10.95, 2.46)  & (15.07, 1.13)  & (18.25, 0.85)  & (21.04, 0.78) &(23.61, 0.81) \\
         Algorithm \ref{alg: SCA-SGPI}, $\Ns=0$     & (10.93, 2.46)  & (15.07, 1.13)  & (18.27, 0.85)  & (21.13, 0.77) &(23.78, 0.78) \\
         LD Algorithm \ref{alg: SCA-SGPI}, $\Ns=3M$     & (10.95, 2.46)  & (15.04, 1.14)  & (18.12, 0.87)  & (20.77, 0.79) &(23.38, 0.80) \\
        \bottomrule
    \end{tabular}
    \vspace{-7mm}
\end{table*}

Table\ \ref{tab:comparison} presents the communications and sensing performance of the considered algorithms for different numbers of transmit antennas. The required average CPU time is also evaluated and shown in Fig.\ \ref{fig:Nt} for the case of Intel(R) Xeon(R) Gold 6226 CPUs. Due to the exponentially increasing computational cost of WMMSE-SDR, its results are provided for only 32 or fewer transmit antennas. Table\ \ref{tab:comparison} demonstrates that the considered algorithms exhibit comparable performance for the same value of $N_{\rm t}$. With more transmit antennas, the communications performance is better at the cost of sensing performance. Note that $\Ns=0$ indicates that the information signals are used for both communications and sensing without dedicated radar signals.  Remarkably, although Algorithm 1 with $\Ns=0$ requires a shorter run time than with $\Ns=3M$, it provides a slight performance gain due its convergence to better locally optimal solutions. Furthermore, Algorithm 1 with $\Ns=3M$, which utilizes the low-dimensional (LD) beamforming structure stated in Proposition \ref{pro:ld structure}, achieves the lowest average CPU time, verifying its efficiency and illustrating its potential for extremely large MIMO systems.

\begin{figure}[t]
\center
\includegraphics[width=0.45\textwidth]{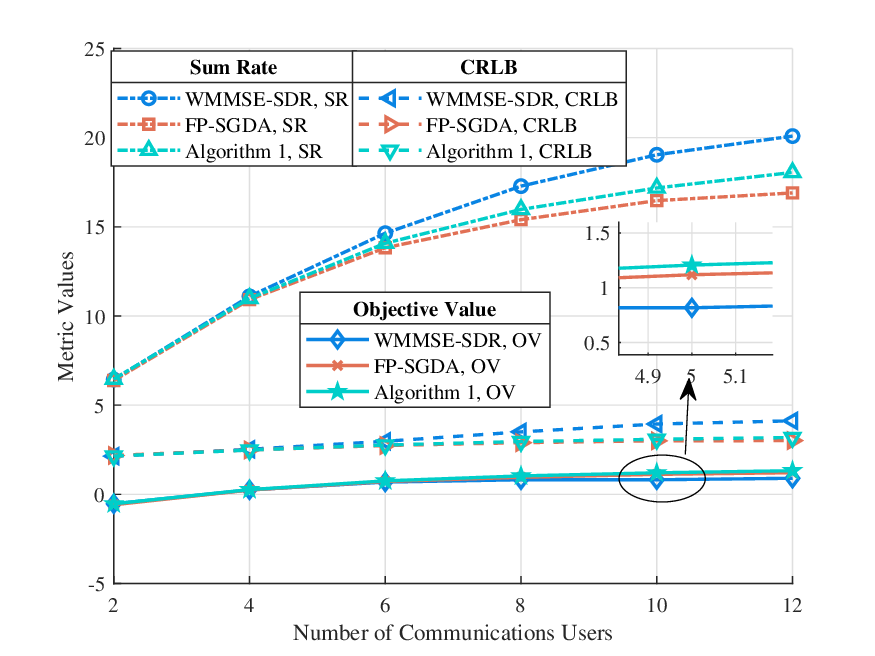}
\vspace{-2mm}
    \caption{Metric values versus the number of communications users.}
    \label{fig:K}
\end{figure}

Fig. \ref{fig:K} plots the value of the communications, sensing, and overall objective function for $K=2,4,\ldots,12$. All three algorithms exhibit a similar trend: a significantly improving communications accompanied by a degradation in the sensing accuracy. This occurs because, as the number of communications users increases, the SR becomes large enough to dominate the ISAC beamforming objective function. The three algorithms show distinct behaviors: the WMMSE-SDR achieves the largest SR but the worst sensing accuracy, FP-SGDA achieves the best sensing accuracy but the worst SR, while the proposed algorithm offers the best overall performance in terms of the objective.

\begin{figure}[t]
\center
\includegraphics[width=0.45\textwidth]{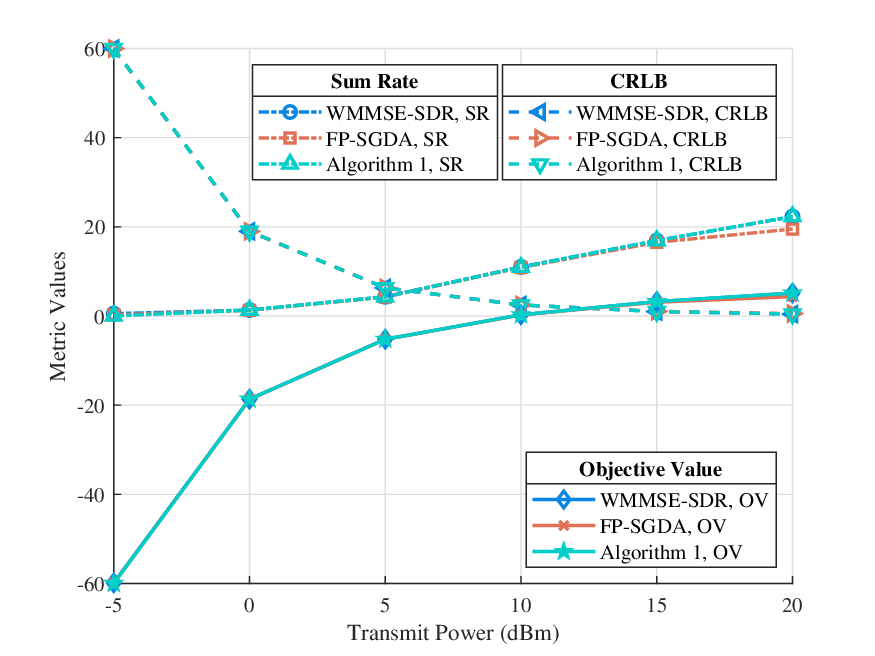}
\vspace{-2mm}
    \caption{Metric values versus the transmit powers.}
    \label{fig:power}
\end{figure}

Fig.\ \ref{fig:power} plots the values of the various objective functions' performances for transmit powers from -5 dBm to 20 dBm. Unlike the previous case, both communications and sensing performances improve as the transmit power increases. All three algorithms exhibit metrics across most regions, except that FP-SGDA experiences a slight performance degradation when the transmit power reaches 20 dBm.

\section{Conclusion}\label{Sec:concu}
We have investigated the fundamental problem of maximizing the weighted sum of the sum rate and the trace of the FIM under a total power constraint for ISAC systems. We first identified the full power consumption property for ISAC systems. Using this property and leveraging lower bounds as surrogates for the intractable objective function, we proposed an efficient low-complexity iterative algorithm that solves the optimization with provable convergence. Furthermore, we obtained the optimal beamforming structure for CRLB-based ISAC systems, enabling a significant reduction in the dimension of the optimized variables to further enhance computational efficiency. Numerical simulations verified the superiority of the proposed design over the considered baselines.

\par The developed algorithms are not restricted to the basic ISAC systems considered in this study; they can be readily extended to advanced ISAC scenarios, such as reconfigurable intelligent surface-aided ISAC and ISAC with near-field communications. Moreover, the proposed full-power property and low-dimensional beamforming structure may also be beneficial for cell-free massive MIMO systems and deep unfolding-based beamforming designs, which we leave for future work.

\appendix
\section{Appendix}
\numberwithin{lemma}{subsection} 
\numberwithin{corollary}{subsection} 
\numberwithin{remark}{subsection} 
\numberwithin{equation}{subsection}	

	\subsection{Proof of Lemma~\ref{lem:FIM}}
 \label{App:derivation of FIM}
	Recall that $\mathbf F$ is the FIM associated with the target parameters $\bm\omega=[\bm\theta,\bm\phi,\Re\{\bm\alpha\},\Im\{\bm\alpha\}]^\T$, and partition $\bm\omega$ as follows: $\bm{\omega}_1 = \bm{\theta}, \bm{\omega}_2 = \bm{\phi}, \bm{\omega}_3 = \Re\{\bm{\alpha}\}, \bm{\omega}_4 = \Im\{\bm{\alpha}\}$. We can then express $\mathbf F$ in \eqref{FIM:original} as a block matrix:
 \begin{align*}
		\mathbf F=  \frac{2}{\sigma_s^2}
		\begin{bmatrix}
			 \mathbf F_{\bm\theta\bm\theta}   & \mathbf F_{\bm\theta\bm\phi}&\mathbf F_{\bm\theta\Re\{\bm\alpha\}}& \mathbf F_{\bm\theta\Im\{\bm\alpha\}}\\
			\mathbf F_{\bm\theta\bm\phi}^\T &\mathbf F_{\bm\phi\bm\phi} & \mathbf F_{\bm\phi\Re\{\bm\alpha\}}& \mathbf F_{\bm\phi\Im\{\bm\alpha\}}\\
    \mathbf F_{\bm\theta\Re\{\bm\alpha\}}^\T & \mathbf F_{\bm\phi\Re\{\bm\alpha\}}^\T & \mathbf F_{\Re\{\bm\alpha\}\Re\{\bm\alpha\}}&\mathbf F_{\Re\{\bm\alpha\}\Im\{\bm\alpha\}}\\
    \mathbf F_{\bm\theta\Im\{\bm\alpha\}}^\T &\mathbf F_{\bm\phi\Im\{\bm\alpha\}}^\T &\mathbf F_{\Re\{\bm\alpha\}\Im\{\bm\alpha\}}^\T &\mathbf F_{\Im\{\bm\alpha\}\Im\{\bm\alpha\}}
		\end{bmatrix},
\end{align*}
where each block is defined as 
\[\mathbf F_{\bm\omega_i\bm\omega_j}=\frac{2}{\sigma_\rms^2}\Re \left\{\sum_{l=1}^L\frac{\partial \mathbf v_{\mathrm{s}}^\H[l]}{\partial \bm\omega_i}\frac{\partial \mathbf v_{\mathrm{s}}[l]}{\partial \bm\omega_j} \right\} ,\forall i,j\in\{1,2,3,4\}.\]
The partial derivative with respect to \(\theta_i\) is given by
\[\frac{\partial\mathbf v_\rms[l]}{\partial\theta_i}=\dot{\mathbf B}_{\bm\theta}\mathbf e_i\mathbf e_i^\T\mathbf U\mathbf A^\H\mathbf x[l]+\mathbf B\mathbf e_i\mathbf e_i^\T\mathbf U\dot{\mathbf A}_{\bm\theta}^\H\mathbf x[l], \forall i\in\{1,\ldots,M\}, \]
where $\mathbf e_i=[0,\ldots,1,\ldots,0]^\T$ denotes a vector with $1$ located at the \(i\)-th position and zeroes elsewhere. Then
\begin{equation*}\label{testasd}
    \begin{aligned}
        \mathbf F_{\theta_i\theta_j}&=\frac{2}{\sigma_\rms^2}\sum_{l=1}^L\Re\left\{\rmtr\left(\frac{\partial \mathbf v_{\mathrm{s}}[l]}{\partial \theta_j} \frac{\partial \mathbf v_{\mathrm{s}}^\H[l]}{\partial \theta_i}  \right)\right\}\\
        &=\frac{2L}{\sigma_\rms^2}\Re\left\{\rmtr\left(\mathbf e_j^\T\mathbf U\mathbf A^\H\mathbf R_{\mathbf x}\mathbf A\mathbf U^\H\mathbf e_i\mathbf e_i^\T\dot{\mathbf B}_{\bm\theta}^\H \dot{\mathbf B}_{\bm\theta}\mathbf e_j \right.\right.\\
        &\quad\quad\,\,\,+\mathbf e_j^\T\mathbf U\mathbf A^\H\mathbf R_{\mathbf x}\dot{\mathbf A}_{\bm\theta}\mathbf U^\H\mathbf e_i\mathbf e_i^\T\mathbf B^\H\dot{\mathbf B}_{\bm\theta}\mathbf e_j\\
        &\quad\quad\,\,\,+ \mathbf e_j^\T\mathbf U\dot{\mathbf A}_{\bm\theta}^\H\mathbf R_{\mathbf x}{\mathbf A}\mathbf U^\H\mathbf e_i\mathbf e_i^\T\dot{\mathbf B}_{\bm\theta}^\H{\mathbf B}\mathbf e_j\\
        &\left.\left.\quad\quad\,\,\,+\mathbf e_j^\T\mathbf U\dot{\mathbf A}_{\bm\theta}^\H\mathbf R_{\mathbf x}\dot{\mathbf A}_{\bm\theta}\mathbf U^\H\mathbf e_i\mathbf e_i{\mathbf B}^\H{\mathbf B}\mathbf e_j\right)\right\} 
    \end{aligned}
\end{equation*}
\begin{equation*}
    \begin{aligned}
        &=\frac{2L}{\sigma_\rms^2}\Re\left\{(\mathbf U\mathbf A^\H\mathbf R_{\mathbf x}\mathbf A\mathbf U^\H)_{ji}(\dot{\mathbf B}_{\bm\theta}^\H\dot{\mathbf B}_{\bm\theta})_{ij}\right.\\
  &\quad\quad\,\,\,+(\mathbf U\mathbf A^\H\mathbf R_{\mathbf x}\dot{\mathbf A}_{\bm\theta}\mathbf U^\H)_{ji}({\mathbf B}^\H\dot{\mathbf B}_{\bm\theta})_{ij} \\
  &\quad\quad\,\,\,+(\mathbf U\dot{\mathbf A}_{\bm\theta}^\H\mathbf R_{\mathbf x}\mathbf A\mathbf U^\H)_{ji}(\dot{\mathbf B}_{\bm\theta}^\H{\mathbf B})_{ij}\\
  &\quad\quad\,\,\,\left.+(\mathbf U\dot{\mathbf A}_{\bm\theta}^\H\mathbf R_{\mathbf x}\dot{\mathbf A}_{\bm\theta}\mathbf U^\H)_{ji}({\mathbf B}^\H{\mathbf B})_{ij}\right\},
    \end{aligned}
\end{equation*}
where $\mathbf X_{ij}$ denotes the \((i,j)\)-th element of $\mathbf X$, $\mathbf R_{\mathbf x}\triangleq\mathbf x[l]\mathbf x^\H[l]=\Rx$, and we have used the property $\rmtr(\mathbf X\mathbf Y\mathbf Z)=\rmtr(\mathbf Y\mathbf Z\mathbf X)$. Hence, we have
\[\mathbf F_{\theta_i\theta_j}=\frac{2L}{\sigma_\rms^2}\Re\{[\mathbf F_{11}]_{ij}\} \quad \Rightarrow \quad \mathbf F_{\bm\theta\bm\theta}=\Re\{\mathbf F_{11}\},\]
with $\mathbf F_{11}$ given in \eqref{FIM:blocks}. The remaining terms in the FIM can be readily calculated in the same way with the following partial derivatives
    \begin{align}
        \frac{\partial\mathbf v_\rms[l]}{\partial\phi_i}&=\dot{\mathbf B}_{\bm\phi}\mathbf e_i\mathbf e_i^\T\mathbf U\mathbf A^\H\mathbf x[l]+\mathbf B\mathbf e_i\mathbf e_i^\T\mathbf U\dot{\mathbf A}_{\bm\phi}^\H\mathbf x[l],\nonumber\\
        \frac{\partial\mathbf v_\rms[l]}{\partial\Re\{\alpha_i\}}&=\mathbf B\mathbf e_i\mathbf e_i^\T\mathbf A^\H\mathbf x[l],\,\,\,\frac{\partial\mathbf v_\rms[l]}{\partial\Im\{\alpha_i\}}=j\mathbf B\mathbf e_i\mathbf e_i^\T\mathbf A^\H\mathbf x[l].\nonumber
    \end{align}

\subsection{Proof of Proposition \ref{pro:full power}}\label{property:full power}

  We prove this proposition by contradiction, assuming that there exists a locally optimal solution $\mathbf W^\diamond$ such that $\|\mathbf W^\diamond\|_\rmf^2<P_{\mathrm t}$. To simplify the notation, we define the total received signal power at user $k$ as the sum of the desired signal power, interference power, and noise power, i.e., $T_k=S_k+D_k+\sigma_{\rmc k}^2$, where
    \begin{align}
        &T_k\triangleq \sum_{j=1}^K |\mathbf h_k^\H\mathbf w_{\rmc k}|^2+\|\mathbf h_k^\H\mathbf W_\rms\|^2_\rmf+\sigma_{\rmc k}^2,\\
       \label{definition Dk} &D_k\triangleq \sum_{j=1,j\neq k}^K |\mathbf h_k^\H\mathbf w_{\rmc j}|^2+\|\mathbf h_k^\H\mathbf W_\rms\|^2_\rmf.\\
        &S_k\triangleq |\mathbf h_k^\H\mathbf w_{\rmc k}|^2.
    \end{align}
Let $\mathbf W^\circ\triangleq \varpi \mathbf W^\diamond$ with $\varpi\triangleq \sqrt {P_{\mathrm t}}/\|\mathbf W^\diamond\|_\rmf>1$. The corresponding SINR for user $k$ satisfies the following inequalities:
    \begin{align}
    &\frac{S_k(\mathbf W^\diamond)}{\sigma_{\rmc k}^2}<\frac{\varpi^2 S_k(\mathbf W^\diamond)}{\sigma_{\rmc k}^2}=\frac{S_k(\mathbf W^\circ)}{\sigma_{\rmc k}^2},\, \text{if}\,  D_k(\mathbf W^\diamond)= 0,\nonumber\\
    &\frac{S_k(\mathbf W^\diamond)}{D_k(\mathbf W^\diamond)+\sigma_{\rmc k}^2}<\frac{\varpi^2 S_k(\mathbf W^\diamond)}{\varpi^2 D_k(\mathbf W^\diamond)+\sigma_{\rmc k}^2}=\frac{S_k(\mathbf W^\circ)}{D_k(\mathbf W^\circ)+\sigma_{\rmc k}^2},\nonumber\\
    &\text{if}\,  D_k(\mathbf W^\diamond)\neq 0,\nonumber
\end{align}
    Since the achievable rate $R_k$ is a monotonically increasing function with respect to its corresponding SINR, these inequalities imply the existence of a point $\mathbf W^\circ=\varpi  \mathbf W_\rms^\diamond$ near $\mathbf W_\rms^\diamond$, with an increased objective value for communications component.
    Similarly, for the sensing component, we have
\begin{equation*}\label{eq:sumpower CRLB}
        \begin{aligned}
            -\mathrm{tr}(\mathbf F^{-1}(\mathbf W_\rms^\circ))=-\varpi^2\mathrm{tr}(\mathbf F^{-1}(\mathbf W_\rms^\diamond))
            >-\mathrm{tr}(\mathbf F^{-1}(\mathbf W_\rms^\diamond)),
        \end{aligned}
    \end{equation*}
  which suggests that the sensing objective value also increases at point $\mathbf W^\circ$. The existence of $\mathbf W^\circ$, at which the overall objective value is strictly larger than that of $\mathbf W^\diamond$, contradicts the assumption that $\mathbf W^\diamond$ is a locally optimal point of problem \eqref{P1}. Thus, for any locally optimal point $\mathbf W_\rms^\diamond$, the corresponding power constraint must hold with equality.  
 
\subsection{Proof of Proposition \ref{pro:compact_form}}\label{App:quadratic problem}
Since \eqref{P1:SCA} consists of both communications and sensing components, the proof is divided into two parts. In the first part, we address the communications component. Specifically, the communications term, $\delta_\rmc\sum_{k=1}^K r_k$, is equal to
\begin{align}\label{obj:communications}
        &\delta_\rmc\sum_{k=1}^K (\log(1+\xi_k)-\xi_k-\beta_k\sigma_{\rmc k}^2)+2\delta_\rmc\Re\{\sum_{k=1}^K\rmtr(\mathbf w_{\rmc k}\eta_k\mathbf h_k^\H)\}  \nonumber\\
        &-\delta_\rmc\sum_{k=1}^K\beta_k\left(  \rmtr(\mathbf W_\rmc\mathbf W_\rmc^\H\mathbf h_k\mathbf h_k^\H)+  \rmtr(\mathbf W_\rms\mathbf W_\rms^\H\mathbf h_k\mathbf h_k^\H) \right)   \nonumber \\
        &=\Bar{C}+2\delta_\rmc \Re\{\rmtr(\mathbf W_\rmc\mathbf \Sigma_1\mathbf H^\H) \}-\delta_\rmc\rmtr(\mathbf W_\rmc\mathbf W_\rmc^\H\mathbf H \mathbf\Sigma_2\mathbf H^\H) \nonumber \\
        &\hspace{4cm}-\delta_\rmc\rmtr(\mathbf W_\rms\mathbf W_\rms^\H\mathbf H \mathbf\Sigma_2\mathbf H^\H),
    \end{align}
where $\Bar{C}$ is a constant, $\mathbf \Sigma_1=\mathrm{diag}(\eta_1,\ldots,\eta_K), \mathbf H=[\mathbf h_1,\ldots,\mathbf h_K]$, and $\mathbf \Sigma_2=\mathrm{diag}(\beta_1,\ldots,\beta_K)$.

For the sensing component, given that $\bm\Phi$ and $\mathbf F$ are symmetric matrices, the sensing term $\rmtr(\mathbf \Phi\mathbf F)$ is equal to $\rmtr(\mathbf \Phi^\T\mathbf F)$, which can be expressed as
    \begin{align}\label{eq:first expanded}
        & \Re\{ \rmtr[\mathbf \Phi_{11}^\T\mathbf F_{11}+2 \mathbf \Phi_{12}^\T\mathbf F_{12}+2(\mathbf \Phi_{13}^\T+j\mathbf \Phi_{13}^\T)\mathbf F_{13}+\bm\Phi_{22}^\T\mathbf F_{22}\nonumber\\
        &+2(\bm\Phi_{23}^\T+j\bm\Phi_{24}^\T)\mathbf F_{23}+(\mathbf \Phi_{33}^\T+\mathbf \Phi_{44}^\T+2j\mathbf \Phi_{34}^\T)\mathbf F_{33}]\}.
    \end{align}
The first term of \eqref{eq:first expanded}, i.e., $\Re\{\rmtr[\mathbf \Phi_{11}^\T\mathbf F_{11}]\}$, can be expanded as
\begin{align}\label{eq:first term}
      \Re\{\rmtr(\mathbf \Phi_{11}^\T&[(\mathbf U\mathbf A^\H\mathbf R_{\mathbf x}\mathbf A\mathbf U^\H)^\T\odot(\dot{\mathbf B}_{\bm\theta}^\H\dot{\mathbf B}_{\bm\theta})\nonumber\\
  &+(\mathbf U\mathbf A^\H\mathbf R_{\mathbf x}\dot{\mathbf A}_{\bm\theta}\mathbf U^\H)^\T\odot({\mathbf B}^\H\dot{\mathbf B}_{\bm\theta}) \nonumber\\
  &+(\mathbf U\dot{\mathbf A}_{\bm\theta}^\H\mathbf R_{\mathbf x}\mathbf A\mathbf U^\H)^\T\odot(\dot{\mathbf B}_{\bm\theta}^\H{\mathbf B})\nonumber\\
  &+(\mathbf U\dot{\mathbf A}_{\bm\theta}^\H\mathbf R_{\mathbf x}\dot{\mathbf A}_{\bm\theta}\mathbf U^\H)^\T\odot({\mathbf B}^\H{\mathbf B}) ]\}\nonumber\\
  =\Re\{\rmtr(\mathbf R_{\mathbf x}&[\mathbf A\mathbf U^\H(\mathbf \Phi_{11}\odot(\dot{\mathbf B}_{\bm\theta}^\H\dot{\mathbf B}_{\bm\theta})) \mathbf U\mathbf A^\H\nonumber\\
  &+\dot{\mathbf A}_{\bm\theta}\mathbf U^\H(\mathbf\Phi_{11}\odot({\mathbf B}^\H\dot{\mathbf B}_{\bm\theta}))  \mathbf U\mathbf A^\H \nonumber\\
  &+\mathbf A\mathbf U^\H(\mathbf\Phi_{11}\odot(\dot{\mathbf B}_{\bm\theta}^\H{\mathbf B}))   \mathbf U\dot{\mathbf A}_{\bm\theta}^\H\nonumber\\
  &+\dot{\mathbf A}_{\bm\theta}\mathbf U^\H  (\mathbf\Phi_{11}\odot({\mathbf B}^\H{\mathbf B}))    \mathbf U\dot{\mathbf A}_{\bm\theta}^\H ]\}\nonumber\\
  \triangleq \Re\{\rmtr(\mathbf R_{\mathbf x}& \mathbf Q_{11})\},
\end{align}
where we have utilized the property
\begin{align}
    \Re&\{\mathrm{tr}(\mathbf \Phi_{11}^\T[(\mathbf U\mathbf A^\H\mathbf R_{\mathbf x}\mathbf A\mathbf U^\H)^\T\odot(\dot{\mathbf B}_{\bm\theta}^\H\dot{\mathbf B}_{\bm\theta}) ])\}\nonumber\\
    &=\Re\{\mathrm{tr}(\mathbf R_{\mathbf x}[ \mathbf A\mathbf U^\H(\mathbf \Phi_{11}\odot (\dot{\mathbf B}_{\bm\theta}^\H\dot{\mathbf B}_{\bm\theta}))\mathbf U\mathbf A^\H ])\}.
\end{align}
The remaining terms can be calculated similarly using the corresponding parameter matrices defined in \eqref{FIM:blocks}. By combining \eqref{obj:communications} and \eqref{eq:first expanded} and omitting constant terms, we obtain \eqref{P1SCA:compact form}.

\subsection{Proof of Proposition \ref{pro:gradient}} \label{app:gradient}
We prove proposition \ref{pro:gradient} separately for the communications and sensing components. To simplify the notation, we define $I_k=D_k+\sigma_{\rmc k}^2$, where $D_k$ is defined in \eqref{definition Dk}.
The gradient of $f_\rmc(\mathbf W)$ is given by

    \begin{align}\label{gradient of fc form1}
        \frac{\partial f_\rmc(\mathbf W)}{\partial \mathbf w_{\rmc k}}&= \frac{\partial R_k}{\partial \mathbf w_{\rmc k}}+\sum_{i=1,i\neq k}^K \frac{\partial R_i}{\partial \mathbf w_{\rmc k}} \nonumber\\
        &=2\left(1+\frac{S_k}{I_k}\right)^{-1}\frac{\mathbf h_k\mathbf h_k^\H\mathbf w_{\rmc k}}{I_k} \nonumber\\
        &\quad-2\sum_{i=1,i\neq k}^K\left(1+\frac{S_i}{I_i}\right)^{-1}\frac{S_i}{I_i^2}\mathbf h_i\mathbf h_i^\H\mathbf w_{\rmc k} \nonumber\\
        &=2\frac{\mathbf h_k\mathbf h_k^\H\mathbf w_{\rmc k}}{I_k}-2\sum_{i=1}^K \frac{S_i}{I_iT_i}\mathbf h_i\mathbf h_i^\H\mathbf w_{\rmc k} \nonumber \\
        &=2\mathbf h_k\eta_k^*-2\sum_{i=1}^K \beta_i\mathbf h_i\mathbf h_i^\H \mathbf w_{\rmc k}, \forall k\in\mathcal K,\\
         \frac{\partial f_\rmc(\mathbf W)}{\partial \mathbf W_{\rms}}
         &\!=\!-2 \sum_{k=1}^K \frac{S_k}{I_kT_k}\mathbf h_k\mathbf h_k^\H\mathbf W_{\rms}\!=\!-2 \sum_{k=1}^K \beta_k \mathbf h_k\mathbf h_k^\H\mathbf W_{\rms}, 
\end{align}

where we have used the identities
\begin{align}
&\left(1+\frac{S_k}{I_k}\right)^{-1}=\frac{I_k}{T_k}=1-\frac{S_k}{T_k},\\
     &\eta_k=\frac{\xi_k}{\mathbf h_k^\H\mathbf w_{\rmc k}}=\frac{(\mathbf h_k^\H\mathbf w_{\rmc k})^*}{I_k},\\
     & \beta_k=\frac{\xi_k}{I_k}=\frac{S_k}{I_kT_k}.
\end{align}
Now, we can rewrite ${\partial f_\rmc(\mathbf W)}/{\partial\mathbf W}$ in the following compact form:
\begin{align}\label{gradient of fc}
    \frac{\partial f_\rmc(\mathbf W)}{\partial\mathbf W}=\frac{2}{\delta_\rmc}\mathbf C_1-2\mathbf H\mathbf\Sigma_2\mathbf H^\H\mathbf W,
\end{align}
which is exactly the same as the gradient of $h_\rmc^{[t]}(\mathbf W)$ by substituting $\mathbf W=\mathbf W^{[t]}$.

For the sensing component, the gradient of $f_\rms(\mathbf W)$ is given by
 \begin{align}\label{gradient of fs form 1}
        \frac{\partial f_\rms(\mathbf W)}{\partial \mathbf W}&=\sum_{m,n=1}^4\sum_{i,j=1}^M \frac{\partial f_\rms(\mathbf W)}{\partial [\mathbf F_{\bm\omega_m\bm\omega_n}]_{ij}}\frac{\partial [\mathbf F_{\bm\omega_m\bm\omega_n}]_{ij}}{\partial \mathbf W}\nonumber\\
        &\overset{\underset{\mathrm{(a)}}{}}{=}\sum_{m,n=1}^4\sum_{i,j=1}^M [\mathbf \Phi_{mn}]_{ij} \frac{\partial [\mathbf F_{\bm\omega_m\bm\omega_n}]_{ij}}{\partial \mathbf W}\overset{\underset{\mathrm{(b)}}{}}{=}2\mathbf Q\mathbf W,
    \end{align}
where $(a)$ follows from $\partial f_\rms(\mathbf W)/\partial \mathbf F=\mathbf F^{-2}=\mathbf \Phi$ and $(b)$ follows from a simplification similar to that in the proof of Appendix \ref{App:quadratic problem}. We take the first term $m=1,n=1 $ as an example to show how to simply \eqref{gradient of fs form 1}:
\begin{align*}
    &\sum_{i,j=1}^M [\mathbf \Phi_{11}]_{ij} \frac{\partial [\mathbf F_{\bm\omega_1\bm\omega_1}]_{ij}}{\partial \mathbf W}=\sum_{i,j=1}^M [\mathbf \Phi_{11}]_{ij} \frac{\partial\mathbf F_{\theta_i\theta_j}}{\partial \mathbf W}\\
&=2\sum_{i,j=1}^M [\mathbf \Phi_{11}]_{ij} [\mathbf A\mathbf U^\H(\mathbf e_i\mathbf e_i^\T(\dot{\mathbf B}_{\bm\theta}^\H\dot{\mathbf B}_{\bm\theta})\mathbf e_j\mathbf e_j^\T) \mathbf U\mathbf A^\H\\
  &\quad+\dot{\mathbf A}_{\bm\theta}\mathbf U^\H(\mathbf e_i\mathbf e_i^\T({\mathbf B}^\H\dot{\mathbf B}_{\bm\theta})\mathbf e_j\mathbf e_j^\T)  \mathbf U\mathbf A^\H \\
  &\quad+\mathbf A\mathbf U^\H(\mathbf e_i\mathbf e_i^\T(\dot{\mathbf B}_{\bm\theta}^\H{\mathbf B})\mathbf e_j\mathbf e_j^\T)   \mathbf U\dot{\mathbf A}_{\bm\theta}^\H\\
  &\quad+\dot{\mathbf A}_{\bm\theta}\mathbf U^\H  (\mathbf e_i\mathbf e_i^\T({\mathbf B}^\H{\mathbf B})\mathbf e_j\mathbf e_j^\T)    \mathbf U\dot{\mathbf A}_{\bm\theta}^\H ]\mathbf W\\
    & =2[\mathbf A\mathbf U^\H(\mathbf \Phi_{11}\odot(\dot{\mathbf B}_{\bm\theta}^\H\dot{\mathbf B}_{\bm\theta})) \mathbf U\mathbf A^\H\\
  &\quad+\dot{\mathbf A}_{\bm\theta}\mathbf U^\H(\mathbf\Phi_{11}\odot({\mathbf B}^\H\dot{\mathbf B}_{\bm\theta}))  \mathbf U\mathbf A^\H \\
  &\quad+\mathbf A\mathbf U^\H(\mathbf\Phi_{11}\odot(\dot{\mathbf B}_{\bm\theta}^\H{\mathbf B}))   \mathbf U\dot{\mathbf A}_{\bm\theta}^\H\\
  &\quad+\dot{\mathbf A}_{\bm\theta}\mathbf U^\H  (\mathbf\Phi_{11}\odot({\mathbf B}^\H{\mathbf B}))    \mathbf U\dot{\mathbf A}_{\bm\theta}^\H]\mathbf W \\
  &=2 \mathbf Q_{11}\mathbf W.
\end{align*}

The remaining terms can be calculated similarly. Note that with $\Re\{\rmtr(\mathbf X)\}=\frac{1}{2}\rmtr(\mathbf X+\mathbf X^\H)$, we can obtain the following equivalent form
\begin{equation}\label{gradient of fs}
    \frac{\partial f_\rms(\mathbf W)}{\partial \mathbf W}=(\mathbf Q+\mathbf Q^\H)\mathbf W,
\end{equation}
which is exactly the same as the gradient of $h_\rms^{[t]}(\mathbf W)$. By summing $\delta_\rmc\times \frac{\partial f_\rmc(\mathbf W)}{\partial\mathbf W}$ and $\delta_\rms\times \frac{\partial f_\rms(\mathbf W)}{\partial \mathbf W}$, we obtain \eqref{gradient of f}.

\subsection{Proof of Proposition \ref{pro:convergence}}\label{APP:convergence}\

Recall the definition of a $\rho$-weakly convex function as stated in \cite{chen2022distributed}:
\begin{definition}\label{def:weakly convex}
    A function $f(\mathbf W)$ is $\rho$-weakly convex if there exists a convex function $g(\mathbf W)$ such that $g(\mathbf W)=f(\mathbf W)+\frac{\rho}{2}\|\mathbf W\|_\rmf^2$. 
\end{definition}
Under the assumption that $f(\mathbf W)$ is $\rho$-weakly convex, we construct the convex function
\[g(\mathbf W)=f(\mathbf W)+\frac{\rho}{2}\|\mathbf W\|_\rmf^2.\]
The full power consumption property, as stated in Proposition~\ref{pro:full power}, implies that $\frac{\rho}{2}\|\mathbf W\|_\rmf^2$ is a constant term. This leads to the following equivalent optimization problem
\begin{equation}\label{P1:boundary2}
 \max_{ \mathbf W\in\mathcal B}\,\, \delta_\rmc\sum_{k=1}^K R_k- \delta_\rms\mathrm{tr}(\mathbf{F}^{-1})+ \frac{\rho}{2}\|\mathbf W\|_\rmf^2.
 \end{equation}
For the new convex objective function, we can derive the following linear lower bound at iteration $t$
\begin{equation}\label{proximal lower bound}
\begin{aligned}
      g(\mathbf W)&\geq g(\mathbf W^{[t]})+\langle\nabla g(\mathbf W)\big|_{\mathbf W=\mathbf W^{[t]}},\mathbf W-\mathbf W^{[t]}\rangle\\
      &=\langle\nabla f(\mathbf W)\big|_{\mathbf W=\mathbf W^{[t]}}+\rho\mathbf W^{[t]},\mathbf W\rangle+ \hat{C},
\end{aligned}
\end{equation}
where $\hat{C}$ is a constant, and  $\langle\cdot,\cdot\rangle$ denotes the inner product in the complex matrix space $\mathbb C^{\Nt\times (K+\Nt)}$.

By maximizing this linear lower bound and projecting the solution onto the boundary of the sphere $\mathcal B$, we derive the PGA update rule as
\begin{equation}
\begin{aligned}
       \mathbf W^{[t+1]}&=\bm\Pi_{\mathcal B}(\nabla f(\mathbf W)\big|_{\mathbf W=\mathbf W^{[t]}}+\rho\mathbf W^{[t]})\\
       &=\bm\Pi_{\mathcal B}\left(\mathbf W^{[t]}+\frac{1}{\rho}\nabla f(\mathbf W)\big|_{\mathbf W=\mathbf W^{[t]}}\right),
\end{aligned}
\end{equation}
which is exactly the same as the update formula in \eqref{projection update} when $\lambda=\frac{\rho}{2}$. Assuming $\lambda\geq \frac{\rho}{2}$, we can establish:
    \begin{align*}
        f(\mathbf W^{[t+1]})&\overset{\underset{\mathrm{(a)}}{}}{=}g(\mathbf W^{[t+1]})-\lambda \|\mathbf W^{[t+1]}\|_\rmf^2\\
        &\overset{\underset{\mathrm{(b)}}{}}{\geq} f(\mathbf W^{[t]})+\langle\nabla g(\mathbf W)\big|_{\mathbf W=\mathbf W^{[t]}},\mathbf W^{[t+1]}-\mathbf W^{[t]}\rangle\\
        &\overset{\underset{\mathrm{(c)}}{}}{\geq} f(\mathbf W^{[t]}) ,
    \end{align*}
  where $(a)$ follows from Definition \ref{def:weakly convex}, $(b)$ follows from the assumption $\lambda\geq \frac{\rho}{2}$ and \eqref{proximal lower bound}, and $(c)$ holds since $\langle\nabla g(\mathbf W)\big|_{\mathbf W=\mathbf W^{[t]}},\mathbf W^{[t+1]}-\mathbf W^{[t]}\rangle\geq 0$ for any convex function $g(\mathbf W)$. Since the objective function is bounded within the closed and compact set $\mathcal B$, the sequence $\{f(\mathbf W^{[t]}\}$ is guaranteed to converge. This completes the proof.

\subsection{Proof of Lemma \ref{pro:rank ws}}\label{app:rank ws}

When $\delta_c=0$, the optimal beamforming structure of $\mathbf W_\rms$ simplifies to
\begin{equation}
    \frac{1}{2}\delta_s\left(\mathbf Q^\diamond+\mathbf Q^{\diamond^\H}\right)\mathbf W_\rms^\diamond=\mu^\diamond\mathbf W_\rms^\diamond,
\end{equation}
which implies that the columns of $\mathbf  W_\rms^\diamond$ are eigenvectors of the matrix $\mathbf Q^\diamond+\mathbf Q^{\diamond^\H}$. Therefore, the rank of $\mathbf  W_\rms^\diamond$ is less than or equal to the rank of $\mathbf Q^\diamond+\mathbf Q^{\diamond^\H} $. Recalling the definition of $\mathbf Q$ given in \eqref{definition of q}, $\mathbf Q+\mathbf Q^\H$ can be expressed as the sum of a set of matrices that can be rearranged as
\begin{equation}
\mathbf Q+\mathbf Q^\H=\mathbf A\mathbf X_1+\dot{\mathbf A}_\theta\mathbf X_2+\dot{\mathbf A}_\phi\mathbf X_3,
\end{equation}
where $\mathbf X_1\in\mathbb C^{ M\times\Nt}, \mathbf X_2\in\mathbb C^{ M\times\Nt},\mathbf X_3\in\mathbb C^{ M\times\Nt}$ are matrices whose specific details are omitted. It is straightforward to verify that 
\begin{equation*}
\begin{split}
        \mathrm{rank}(\mathbf Q+\mathbf Q^\H)&\leq \mathrm{rank}(\mathbf A\mathbf X_1)+\mathrm{rank}(\dot{\mathbf A}_\theta\mathbf X_2)+\mathrm{rank}(\dot{\mathbf A}_\phi\mathbf X_3)\\
        &\leq 3M,
\end{split}
\end{equation*}
which completes this proof.

\subsection{Proof of Proposition \ref{pro:ld structure}}\label{app:ld structure}

Similar to the proof of Proposition \ref{pro:gradient}, we prove this proposition for the communications and sensing separately. For the communications subproblem, the optimal beamforming structure of $\mathbf W_\rmc$ reduces to
\begin{equation}\label{eq:ld for c}
\begin{aligned}
    \mathbf W_\rmc^\diamond&=(\mu^\diamond\mathbf I+\mathbf H\mathbf \Sigma_2^\diamond\mathbf H^\H)^{-1}\mathbf H\mathbf \Sigma_1^{\diamond^\H}\\
    &=\mathbf H(\mu^\diamond\mathbf I+\mathbf \Sigma_2^\diamond\mathbf H^\H\mathbf H)^{-1}\mathbf \Sigma_1^{\diamond^\H}\triangleq \mathbf H\mathbf P^\diamond_\rmc,
\end{aligned}
\end{equation}
which indicates that the optimal solution of the SR problem lies in the range space of the channel matrix $\mathbf H$. For the sensing subproblem, the columns of the optimal solution $\mathbf W_\rms^\diamond$ are eigenvectors of the matrix $\mathbf Q^\diamond+\mathbf Q^{\diamond^\H} $. Consequently $\mathbf W_\rms^\diamond$ lies in the range space of matrix $\mathbf Q^\diamond+\mathbf Q^{\diamond^\H} $, i.e.,
\begin{equation}\label{eq:ld fora s}
    \begin{aligned}
        \mathbf W_\rms^\diamond&=\left(\mathbf Q^\diamond+\mathbf Q^{\diamond^\H} \right)\widehat{\mathbf P}_\rms\\
        &=[\mathbf A, \dot{\mathbf A}_\theta, \dot{\mathbf A}_\phi] [\mathbf X_1\widehat{\mathbf P}_\rms, \mathbf X_1\widehat{\mathbf P}_\rms,\mathbf X_1\widehat{\mathbf P}_\rms]^\T\\
        &\triangleq[\mathbf A, \dot{\mathbf A}_\theta, \dot{\mathbf A}_\phi]\mathbf P_\rms^\diamond,
    \end{aligned}
\end{equation}
which implies that $\mathbf W_\rms^\diamond$ lies in the range space of matrix $[\mathbf A, \dot{\mathbf A}_\theta, \dot{\mathbf A}_\phi]$. By introducing new variables $\mathbf P_\rmc$ and $\mathbf P_\rms$ and substituting \eqref{eq:ld for c} and \eqref{eq:ld fora s} back into the original problem \eqref{P1}, we obtain the low-dimensional reformulation \eqref{P1:ld}.

	\bibliographystyle{IEEEbib} 
	\bibliography{IEEEabrv,reference}
%
%

\end{document}